\documentclass[12pt]{article}
\pdfoutput=1
\usepackage{amsfonts}
\usepackage{mathptmx}
\usepackage{amsmath}
\usepackage[top=2cm,left=1cm,right=1cm,bottom=2cm,includehead,includefoot]{geometry}
\usepackage{cite}
\usepackage{color}
\usepackage{setspace}
\usepackage[T1]{fontenc}
\usepackage[latin1]{inputenc}
\usepackage{verbatim}
\usepackage{subfigure}
\usepackage{float}
\usepackage[pdftex]{graphicx}

\begin{document}

\title{Observational constraints of the gravitational waves in the Brans-Dicke theory: Einstein frame and Jordan-Brans-Dicke frame}
\author{R. C. Freitas\thanks{e-mail: \texttt{rc\_freitas@terra.com.br }} \ and S. V. B. Gon\c{c}alves\thanks{e-mail: \texttt{sergio.vitorino@pq.cnpq.br}} \\
\mbox{\small Universidade Federal do Esp\'{\i}rito Santo, Centro de Ci\^{e}ncias Exatas, Departamento de F\'{\i}sica}\\
\mbox{\small Av. Fernando Ferrari, 514 - Campus de Goiabeiras, CEP
29075-910, Vit\'oria, Esp\'{\i}rito Santo, Brazil}}
\date{\today}
\maketitle

\begin{abstract}
         We investigate the quantum origin of the primordial cosmological gravitational waves in the Brans-Dicke theory in the two conformally related
         frames, the Jordan-Brans-Dicke frame and the Einstein frame. We calculated the theoretical observable in both frames and we
         compared both with General Relativity. We compute the number of gravitons $N_k$ produced during inflation and the observables: power spectrum
         $P_T$, spectral index $n_T$ and energy density $\Omega_k$. The comparison shows that for the case of the particles number $N_k$ the results are the same
         in both frames and in General Relativity when the Brans-Dicke parameter is much bigger than unity. For the spectral index $n_T$ we show that
         it is possible to get a scale invariant perturbation in the Jordan-Brans-Dicke frame when $\omega\rightarrow\infty$ and in the Einstein frame when
         $\omega\rightarrow\pm\infty$. In both frames, the results found for the power spectrum $P_T$ and the energy density $\Omega$ show that the
         preferred values of $\omega$ are differente from that that are found in the local tests.

\vspace{0.7cm}

\par
KEYWORDS: scalar-tensor gravity, gravitational waves, conformal transformations
\par
\vspace{0.7cm}
\par
PACS numbers: 04.30.-w, 98.80.-K

\end{abstract}

\section{Introduction}
\label{intro}
\par
The gravitational waves are tensorial fluctuations in the metric of spacetime. This particular perturbation is not explicitly coupled with the energy density and pressure of the matter of the Universe and does not contribute to the gravitational instability that generates the cosmological structures we see today. On the other hand this study is of great interest because it supplies the specific signature of the metric theory of gravity.
There is a special interest in the primordial gravitational waves of quantum origin because they may have left special signatures in the polarization of the Cosmic Microwave Background (CMB) anisotropies \cite{paul1,paul2, mass}. Moreover, since the gravitational waves were generated in the early Universe, before the time of last scattering of CMB photons, they can be a window to the primordial phase of evolution of the Universe
and can help testing the initial conditions of the scalar field. These waves are predicted by Einstein's theory of General Relativity (GR), but they still have to be directly detected. Great efforts have been done in this sense and there is hope that a new generation of experiments in space may allow this detection \cite{gwobs}. The gravitational waves spectrum frequency extends over a wide range of interest, from $10^{-18}Hz$ to $10^{8}Hz$, depending of the sources that generate those waves. Many works have been done in order to identify specific signatures of cosmological models in the spectra of gravitational waves, for instance, in the case of quintessence model \cite{uzan} and string cosmology \cite{ungarelli,sanchez,gasperini}.
\par
In 1999, the SN Ia observations \cite{riess} showed that the Universe is currently undergoing accelerated expansion. A possible theoretical explanation for this acceleration is the vacuum energy with negative pressure, called dark energy (that violates the strong energy condition). There are many dark energy models, for example quintessence \cite{quinte}, k-essence \cite{k}, phantom \cite{fantasma} and Chaplygin gas \cite{chap}. The simplest candidate for the dark energy is the cosmological constant $\Lambda$ whose equation of state is $p = -\rho$. But there are some problems with this alternative as, e.g., why the cosmological constant is so small, nonzero, and comparable to the critical density at the present? So, it is interesting to study dark energy models as a possible theory of k-essence coupled to gravity since it involves the simplest form of nonlinear kinetic term for the scalar field $\phi$ \cite{rose}.
\par
The existence of a classical scalar field in the nature has been considered in many alternative theories of gravitation. Scalar-tensor gravity theories are a particular case of the huge class called Extended Theories of Gravity which are revealing a useful paradigm to deal with several problems in cosmology and related areas including fundamental physics \cite{ext}. The prototype of this kind of theories is the Brans-Dicke theory (BDT) \cite {1, 2, 3, 4} and its revival leads us to ask if the generation of gravity waves can be modified through the introduction of this scalar field. Here we propose to study the evolution of tensorial fluctuation in the traditional BDT and its quantum origin and calculate the observables that can be compared with the future observational data. The main interest in the presence of the scalar field is the possibility that it can change the evolution of gravity waves in comparison with the results obtained with GR \cite{barrow}. But the situation is not so simple: there are other problems related with the BDT as the reference frame problem.
\par
As others scalar-tensor theories, the BDT can be formulated in the Einstein frame (EF) or in the Jordan-Brans-Dicke frame (JBDF), which are conformally related.  We have two conformally related frames, $\bar{{ds}^2}$ and $ds^2$, if we multiply the line element $ds^2$ by a nonvanishing function of spacetime coordinates $\Omega^2 (x^{\mu})$ and the resulting line element have the same light cone that the first one. Thus, the purpose of the conformal transformation is to give another representation of the theory that is equivalent to it. This can be made by a simple transformation of units \cite{dicke}. In the case of the BDT we are looking for a system of units where the gravitational coupling is constant.
\par
In the specific case of the BDT which conformal frame is the physical one is a very contentious issue \cite{faraoni}. The most common argument against the JBDF found in the literature is that the scalar field energy density can assume negative values. This happens because of the nonmininal coupling between the scalar field and geometry. The terms with the second covariant derivatives of the scalar field in the Brans-Dicke field equations contain the connection $\Gamma^{\alpha}_{\mu\nu}$ and therefore part of the dynamical description of gravity. But the energy density can be made nonnegative with the help of a new connection $\tilde{\Gamma}^{\alpha}_{\mu\nu}$ \cite{santiago} which is also associated with the physical metric $g_{\mu\nu}$.
\par
By a conformal rescaling of the metric $\bar g_{\mu\nu} = \bar g_{\mu\nu}(g_{\mu\nu}, \phi)$  and a nonlinear field redefinition $\bar\phi = \bar\phi(g_{\mu\nu}, \phi)$ we obtain the EF of the BDT, that has a positive definite energy. In this frame, however, there is a violation of the equivalence principle, due to an anomalous coupling of the scalar field $\phi$ to ordinary matter. This violation is small and compatible with the available tests of the equivalence principle. It is, indeed, a low- energy manifestation of general compactified theories where, e.g., the dilaton in the n-dimensional Kaluza-Klein theory plays the role of the Brans-Dicke scalar field $\phi$.
\par
Classically both the theories are equivalent and describe the same physical system with different names. The equivalence appears as follows: the space of solutions in the JBDF is isomorphic to that the EF and the isomorphism is given by the relations $g_{\mu\nu}\rightarrow \bar g_{\mu\nu}$ and $\phi\rightarrow\bar\phi$. But from the physical point of view the situation is very different mainly due the in definition concerning the energy and related to the coupling between the scalar field and the gravitation. Moreover, another factor that distinguishes the two frames from the classical view is that in the case of EF gravitational radiation is generated by the quadrupole term while in the case of JBDF we also find the terms of monopole and dipole radiation \cite{scharre}. Nevertheless it has been a tradition to consider the Jordan metric as physical in the BDT, since the initial proposal of Brans-Dicke. On the other hand, a fundamental assumption in the theory of gravitation is that the gravitational interaction is generated by the massless spin-two graviton. Once one accepts this hypothesis, one must accept the EF as the physical spacetime metric in the BDT. Therefore, so as we can see, this is an unsolved problem related to the BDT. To shed light on this problem, far from theoretical issues, we calculate the physical observables related to the BDT through the quantization of gravitational waves and we hope that the observations decide what is the best answer to this important question.
\par
This Letter is organized as follows. In section 2 we study the background solutions, the gravitational waves equation, the quantum gravitational waves equation and the physical observables in the JBDF. In section 3 we make the same analysis but now in the EF. Finally, in section 4 we discuss the results and show our conclusions.
\par
We use in this Letter standard tensor notation: the greek indices run from zero to three; the latin indices run from one to three; the spacetime metric is taken to have signature $+2$, or $(-,+,+,+)$; covariant derivatives are indicated by semicolons; partial derivatives are indicated by commas and the scalar field $\phi$ is a time function. Moreover, we work using units in which $\hbar = c = k_B = 1$.

\section{Jordan-Brans-Dicke frame}
\label{jordan}
\subsection{Background solutions}
\label{subsectionbackJ}
In this scenario, the action of the BDT is given by  \cite{plinio}
\begin{equation}
\label{actionJF}
\mathcal{S} = \int d^4 x\sqrt{-g}\biggl[\phi R - \omega\biggl(\frac{g^{\mu\nu}\phi,_{\mu}\phi,_{\nu}}{\phi}\biggr) + 16\pi\mathcal{L}_{mat}\biggr]\quad.
\end{equation}
\par
We assume the background of the Universe as being spatially flat (k = 0), homogeneous and isotropic, {\it i.e.}, it is described by the flat Friedmann-Lema\^itre-Robertson-Walker (FLRW) metric
\begin{equation}
               \label{metric}
               \mathrm{d}s^2 = -\mathrm{d}t^2 + a^{2}(t)\left(\mathrm{d}x^2+\mathrm{d}y^2+\mathrm{d}z^2 \right)
               \hspace{0.2cm} \textrm{,}
\end{equation}
where $a(t$) is the scale factor of the Universe. This choice is because the present observations indicate that the density parameter $\Omega$ is close to unity \cite{Omega}. With the metric in this form we are considering a state on a certain three-dimensional space-like surface in spacetime, which is due to the topological structure of the FLRW spacetime - the usual canonical ADM decomposition of the metric.
\par
Considering the matter content of the Universe as a barotropic fluid, $p = \alpha\rho$, we obtain the equations of motion
\begin{eqnarray}
\label{motion1}
- 3\frac{\ddot a}{a} &=& \frac{8\pi}{\phi}\rho\biggl(\frac{2 + \omega + 3\alpha + 3\alpha\omega}{3 + 2\omega}\biggr) + \omega\frac{\dot\phi^2}{\phi^2} + \frac{\ddot\phi}{\phi}\quad, \\
\label{motion2}
\ddot\phi + 3\frac{\dot a}{a}\dot\phi &=& \frac{8\pi}{3 + 2\omega}\rho(1 - 3\alpha)\quad.
\end{eqnarray}
\par
The above equations must be supplemented by a conservation equation of the energy-momentum tensor
\begin{equation}
\label{energia}
\dot\rho + 3\frac{\dot a}{a}\rho(1 + \alpha) = 0\quad.
\end{equation}
\par
Background solutions can be obtained assuming that the scale factor $a(t)$ and the scalar field $\phi(t)$ have a power-law form
\begin{equation}
a(t)\propto t^r~~~~ ,~~~~\phi(t)\propto t^s\quad.
\end{equation}
\par
In what follows we shall consider the inflation case ($\alpha = -1$). In this scenario we have a drastic expansion of the Universe during the early period of the Big Bang. The background solutions are
\begin{equation}
\label{infla}
s = 2~~, ~~~~ r = \omega + \frac{1}{2}\quad.
\end{equation}

\subsection{The gravitational waves equation}
\label{subsectionwavesJ}
Cosmological gravitational waves are obtained by means of a small correction $h_{ij}$ in equation (\ref{metric}), which represents the metric. Hence, the general expression of the metric (\ref{metric}), related to the unperturbed metric, is replaced by
\begin{equation}
               \label{metric1}
               \mathrm{d}s^2 = -\mathrm{d}t^2 + \left[a^{2}(t)\delta_{ij}+h_{ij}\right]\mathrm{d}x^i \mathrm{d}x^j
               \hspace{0.2cm} \textrm{.}
\end{equation}
We use the above metric in the field equation to obtain the resulting gravitational waves equation
\begin{equation}
\label{gw1}
               \ddot{h}_{ij}+\left(\frac{\dot{\phi}}{\phi} -\frac{\dot{a}}{a}\right)\dot{h}_{ij}+
               \left[\frac{k^2}{a^2}-2\frac{\ddot{a}}{a} -2\frac{\dot{a}\dot{\phi}}{a\phi} \right]h_{ij}=0
               \hspace{0.2cm} \textrm{,}
\end{equation}
where $k$ is the wavenumber, the dots indicate cosmic time derivatives and we have written $h_{ij}(t,\vec x) =  h(t)Q_{ij}$, where $Q_{ij}$ are the eigenmodes of the Laplacian operator, such that $Q_{ii} = Q_{ki,k} = 0$.
\par
Performing the transformation from the cosmic time to the conformal time, $a(t)~d\eta = dt$, and representing the derivatives with respect to $\eta$ by primes, equation (\ref{gw1}) assumes the form
\begin{equation}
\label{gw2}
h_{ij}''+\left(\frac{\phi'}{\phi}-2\frac{a'}{a} \right)h_{ij}'+  \left[k^2-2\frac{a''}{a}+2\frac{a'^2}{a^2}-2\frac{a'\phi'}{a\phi} \right]h_{ij}=0
\hspace{0.3cm} \textrm{.}
\end{equation}
\par
To resolve this differential equation we need to perform the Fourier transformation
\begin{equation}
               \label{eq:perturbfourier}
h_{ij}(\vec{x},\eta)=\sqrt{16\pi}\sum_{\lambda=\otimes,\oplus}\int{\frac{\mathrm{d}^{3}k}{(2\pi)^{3/2}}
               \epsilon_{ij}^{(\lambda)}(\hat{k})\frac{a(\eta)\mu_{(\lambda)}(\eta)}{\sqrt{\phi(\eta)}}e^{-\dot{\imath}\vec{k}\cdot\vec{x}} }
               \hspace{0.3cm} \textrm{,}
\end{equation}
where the polarization tensor $\epsilon_{ij}^{(\lambda)}(\hat{k})$ can be decomposed as $\epsilon_{ij}^{(\lambda)}(\hat{k})\epsilon^{ij(\lambda')}(\hat{p})=2\delta^{\lambda \lambda'}\delta^{(3)}(\vec{k}-\vec{p})$
to the two polarization states $\otimes$ e $\oplus$. Then, we have
\begin{equation}
               \label{eq:mudif1}
\mu_{(\lambda)}''(\eta) + \biggl[k^2 + \frac{a''}{a}+\frac{1}{2}\frac{\phi''}{\phi} - \frac{1}{4}\left(\frac{\phi'}{\phi}\right)^{2}+\frac{a'\phi'}{a\phi}\biggl]\mu_{(\lambda)}(\eta)=0
               \hspace{0.2cm} \textrm{.}
\end{equation}
In this way, we rewrote the equation (\ref{gw2}) in terms of a harmonic oscillator.
\par
We can express the equation (\ref{eq:mudif1}) in terms of the new parameter $\Phi(\eta)\equiv a(\eta)\sqrt{\phi(\eta)}$. Hence, the equation (\ref{eq:mudif1}) becomes
            \begin{equation}
               \label{eq:mudif}
               \mu_{(\lambda)}''(\eta)+\left[k^2 -\frac{\Phi''(\eta)}{\Phi(\eta)} \right]\mu_{(\lambda)}(\eta)=0 
               \hspace{0.3cm} \textrm{.}
            \end{equation}   
         \par
With the background solution of the scale factor and the scalar field and considering $\alpha = -1$ we get
\begin{equation}
\label{eq:potencial}
\frac{\Phi''(\eta)}{\Phi(\eta)}=\frac{2(3+2\omega)(1+2\omega)}{(1-2\omega)^2~|\eta|^2}~~\mbox{,}\quad \mbox{with} \quad  -\infty  < \eta \leq -\eta_{1}\quad.
\end{equation}
\par
The solution in the inflationary era is
\begin{equation}
                \label{eq:solinflacao}
                \mu(k\eta)=\sqrt{\eta}\left(A_{1}\mathcal{H}^{(1)}_{\nu}(k|\eta|)+A_{2}\mathcal{H}^{(2)}_{\nu}(k|\eta|) \right)
                \hspace{0.1cm} \textrm{,}
             \end{equation}
where $A_{1}$ and $A_{2}$ are integration constants, $\mathcal{H}^{(1)}$ and $\mathcal{H}^{(2)}$ are the Hankel functions of first and second kind, respectively, and $\nu=\frac{5+6\omega}{2(1-2\omega)}$ is the order of the Hankel functions. 

\subsection{The quantum gravitational waves equation}
\label{subsectionquantumJ}

Since the primordial gravitational waves are of quantum origin we need to quantize the perturbation $h_{ij}$. In order to do that we need to put the equations for the perturbation into the Hamiltonian form. Dirac \cite{dirac} showed that it is possible to make the quantization without the problems  resulting  from the diffeomorphism  invariance. In order to proceed with the quantization we start with the Lagrangian density of the gravitational waves in the BDT, given by
\begin{equation}
\label{lagran01}
                  \mathcal{L}=\frac{\phi}{16\pi} g^{\mu\nu}\partial_{\mu}h_{ij}\partial_{\nu}h^{ij}
                  \hspace{0.2cm} \textrm{,}
\end{equation}
that leads to two corresponding Hamiltonians
\begin{eqnarray}
                 \label{eq:hamilt1}
                 H^{(1)}&=&-\frac{1}{2}\int{\mathrm{d}^{3}x\left[\tilde{\pi}^2 -2\frac{\Phi'}{\Phi}\mu\tilde{\pi}+
                 \delta^{ij}\mu_{,i}\mu_{,j}\right]} \quad, \\
                 \label{eq:hamilt2}
                 H^{(2)}&=&-\frac{1}{2}\int{\mathrm{d}^{3}x\left[\pi^2 -\frac{\Phi''}{\Phi}\mu^2 +
                 \delta^{ij}\mu_{,i}\mu_{,j}\right]} 
                 \hspace{0.2cm} \textrm{,}
\end{eqnarray}
where $\pi$ and $\tilde{\pi}$ are the canonical momenta. The above Hamiltonians differ each other by addiction or subtraction of a total conformal time derivative.
\par
The expression of the quantum Hamiltonian obtained by the equation (\ref{eq:hamilt2}) for the gravitational waves in the BDT is constructed by using the canonical commutation relations
\begin{equation}
                 \label{eq:comutacao}
                 \left[\hat{\mu}(\vec{x},\eta),\hat{\pi}(\vec{y},\eta) \right]=\dot{\imath}\delta^{(3)}(\vec{x}-\vec{y})
                 \hspace{0.2cm} \textrm{.}
\end{equation}
\par
So, by transforming the normal modes of oscillation of the gravitational waves in field and momentum operators in the Heisenberg description of quantum mechanics we obtain the Hamiltonian
\begin{equation}
                 \label{eq:hamiltoniana2}
                 H^{(2)}\rightarrow \hat{H}^{(2)}=-\frac{1}{2}\int{\mathrm{d}^{3}x\left[\hat{\pi}^2 
                 -\frac{\Phi''}{\Phi}\hat{\mu}^2 +\delta^{ij}\hat{\mu}_{,i}\hat{\mu}_{,j}\right]} 
                 \hspace{0.2cm} \textrm{,}
              \end{equation}
where the field operator is constructed from its classical field as follows
\begin{equation}
\label{oper01}
                 h_{ij}\rightarrow\hat{h}_{ij}(\vec{x},\eta) = \frac{\sqrt{16\pi}}{a(\eta)\sqrt{\phi(\eta)}}\sum_{\lambda=\otimes,\oplus}\frac{1}{2}
                    \int \frac{\epsilon_{ij}^{(\lambda)}(\vec{k})\mathrm{d}^3k}{(2\pi)^{3/2}}
                    \left(\hat{\mu}_{\vec{k},(\lambda)}
                    e^{-\dot{\imath}\vec{k}\cdot\vec{x}}
                    +\hat{\mu}_{\vec{k},(\lambda)}^{\dagger}e^{\dot{\imath}\vec{k}\cdot\vec{x}}\right)
                    \hspace{0.2cm} \textrm{,}
\end{equation}
with the operator $\hat{\mu}_{\vec{k}}$ given by
              \begin{equation}
                 \hat{\mu}_{\vec{k}}=\hat{a}_{\vec{k}}(\eta_0)f_{k}(\eta)+\hat{a}_{-\vec{k}}^{\dagger}(\eta_0)f_{k}^{*}(\eta)
                 \hspace{0.2cm} \textrm{,}
              \end{equation}
and the function $f_{k}$ satisfying the following equation 
              \begin{equation}
                 f_{k}''+\left(k^{2}-\frac{\Phi''}{\Phi}\right)f_{k}=0 \hspace{0.2cm} \textrm{,}
              \end{equation}
which is the same differential equation that describes the behavior of the classical function $\mu(\eta)$.
\par
Now, to obtain the coherent states of the relic gravitons in the BDT we consider the creation and annihilation operators
\begin{equation}
                 \hat{a}_{\vec{k}}=\sqrt{\frac{k}{2}}\left(\hat{\mu}_{\vec{k}}+\frac{\dot{\imath}}{k}\hat{\pi}_{\vec{k}}\right)
                 \hspace{0.5cm} \textrm{,} \hspace{0.5cm}
                 \hat{a}_{-\vec{k}}=\sqrt{\frac{k}{2}}\left(\hat{\mu}_{\vec{k}}-\frac{\dot{\imath}}{k}\hat{\pi}_{\vec{k}}\right)
                 \hspace{0.2cm} \textrm{,}
              \end{equation}
whose temporal evolution is given by the Heisenberg equation and by the Hamiltonian (\ref{eq:hamilt1}), with the general solutions given by
\begin{eqnarray}
                 \label{eq:solgeralcoefbogo}
                 \hat{a}_{\vec{k}}(\eta)&=&u_{k}(\eta)\hat{a}_{\vec{k}}(\eta_{0})
                 +v_{k}(\eta)\hat{a}_{-\vec{k}}^{\dagger}(\eta_{0}) \hspace{0.2cm} \quad,\nonumber \\
                 \hat{a}_{-\vec{k}}^{\dagger}(\eta)&=&v_{k}^{*}(\eta)\hat{a}_{\vec{k}}(\eta_{0})
                 +u_{k}^{*}(\eta)\hat{a}_{-\vec{k}}^{\dagger}(\eta_{0}) \hspace{0.5cm} \textrm{,}
\end{eqnarray}
where $\eta_{0}$ is a fixed initial time and the functions $u_{k}(\eta)$ and $v_{k}(\eta)$ behave as
              \begin{equation}
                 \label{eq:transfbogo}
                 \frac{\mathrm{d}u_{k}}{\mathrm{d}\eta}=\dot{\imath}ku_{k}+\frac{\Phi'}{\Phi}v_{k}^{*} 
                 \hspace{0.5cm} \textrm{,} \hspace{0.5cm}
                 \frac{\mathrm{d}v_{k}}{\mathrm{d}\eta}=\dot{\imath}kv_{k}+\frac{\Phi'}{\Phi}u_{k}^{*} 
                 \hspace{0.5cm} \textrm{.}
              \end{equation} 
 \par
The equations (\ref{eq:solgeralcoefbogo}) are the Bogoliubov transformations \cite{birrel} of the gravitational waves in the BDT and $u_{k}(\eta)$ and $v_{k}(\eta)$ are the Bogoliubov coefficients that satisfy the relation
              \begin{equation}
                 \left|u_{k}\right|^2-\left|v_{k}\right|^2=1 \hspace{0.3cm} \textrm{,}
              \end{equation}
result that guarantees the unity of the temporal evolution of these operators.
\par
The number operator is defined as $N_k = v_{k}^{*}v_{k}$. So, in our model the number of gravitons is given by            
\begin{equation}
N_k = \left|v_{k}\right|^2=\frac{|f_{k}|^2}{2k}\left(k^2+\left(\frac{\Phi'}{\Phi}\right)^2\right)+  \frac{1}{2k}\left|\frac{\mathrm{d}f_{k}}{\mathrm{d}\eta}\right|^2-            \frac{1}{2k}\frac{\Phi'}{\Phi}\left(f_{k}\frac{\mathrm{d}f^{*}_{k}}{\mathrm{d}\eta}+
f^{*}_{k}\frac{\mathrm{d}f_{k}}{\mathrm{d}\eta}\right) -\frac{1}{2}
\hspace{0.2cm} \textrm{,}
\end{equation}
where the relation between $\nu_k$ and $f_k$ is
\begin{equation}
                 v_{k}^{*}=\frac{\dot{\imath}}{\sqrt{2k}}\frac{\mathrm{d}f_{k}}{\mathrm{d}\eta}+
                           \frac{f_{k}}{\sqrt{2k}}\left(k-\dot{\imath}\frac{\Phi'}{\Phi}\right)
                           \hspace{0.2cm} \textrm{.}
\end{equation}
\par
When $k\eta>>1$ the number operator simplifies to
              \begin{equation}
                 N_k \approx k|f_{k}|^2\left(1+\frac{1}{2k^2}\left(\frac{\Phi'}{\Phi}\right)^2\right)-\frac{1}{2}
                 \hspace{0.3cm} \textrm{.}
              \end{equation}
\par
If $k\eta<<1$ we have
              \begin{equation}
                 N_k \approx k|f_{k}|^2-\frac{1}{2}
                 \hspace{0.3cm} \textrm{.}
              \end{equation}
\par
In Figures~\ref{num01}-\ref{num02} we can see the behavior of the number of particles in the inflation era in the BDT in the case of the JBDF to different values of $\omega$ and small and big values of the frequency and we compare with the EF and with the GR.

\subsection{Physical Observables}
\label{subsectionobsJ}
\par
In this section we apply the formulation of quantum gravitational waves obtained in BDT in the JBDF in order to determine the most common and interesting observables related to the gravitational waves. The observational data in the present do not allow us to include or exclude the presence of a primordial spectrum of relic gravitons compatible with any model and any frame. The idea is that these theoretical results calculated here might in the future be compared with the observations obtained by the projects related to the search for gravitational waves of cosmological origin.

\subsubsection{Power Spectrum}

\par
The power spectrum is related with the two-point correlation function of the tensor modes in the following way \cite{inflacao}
\begin{equation}
                 \label{eq:funccorrelacao}
                 \left\langle0\right|\hat{h}_{ij}^{\dagger}(\vec{x},\eta)\hat{h}^{ij}(\vec{y},\eta)\left|0\right\rangle=
                 \int{P_{T}(k,\eta)\frac{\sin{(kr)}}{kr} \mathrm{d}(\ln{k})}
                 \hspace{0.2cm} \textrm{,}
\end{equation}
where $P_{T}(k,\eta)$ is the power spectrum and the state $\left|0\right\rangle$ is annihilated by $\hat{a}_{\vec{k}}$ .
         \par   
This operator is given by the equation (\ref{oper01}) where the vacuum state $\left|0\right\rangle$, in which there are no particles, is annihilated by the annihilation operator
              \begin{equation}
                 \hat{a}_{-\vec{k}}\left|0\right\rangle=0
                 \hspace{0.3cm} \textrm{.}
              \end{equation}  
         \par
The calculations of the expectation value of the field operators yield
\begin{equation}
                 \left\langle0\right|\hat{h}_{ij}^{\dagger}(\vec{x},\eta)\hat{h}^{ij}(\vec{y},\eta)\left|0\right\rangle=
                    64\pi\int{\frac{\mathrm{d}^3k}{(2\pi)^{3}} \frac{\left|f_{k}(\eta)\right|^2}{a^2\phi}
                     e^{-\dot{\imath}\vec{k}\cdot(\vec{x}-\vec{y})} } 
                     \hspace{0.2cm} \textrm{.}
\end{equation}
         \par
Knowing that gravitons can be produced in isotropic models and using spherical coordinates, we have
\begin{equation}
                 \left\langle0\right|\hat{h}_{ij}^{\dagger}(\vec{x},\eta)\hat{h}^{ij}(\vec{y},\eta)\left|0\right\rangle=
                    64\pi\int{\frac{k^2}{(2\pi)^{2}} \frac{\left|f_{k}\right|^2}{a^2\phi} 
                      \frac{2\sin{(kr)}}{kr} \mathrm{d}k}
                      \hspace{0.2cm} \textrm{.}
\end{equation}
\par
Comparing the equation (\ref{eq:funccorrelacao}) with the previous equation we have the power spectrum of the tensor modes, within the JBDF, parametrized as
              \begin{equation}
                 P_{T}(k,\eta)=32\pi\frac{k^{3}}{\pi^2}\frac{\left|f_{k}\right|^2}{a^2\phi}
                 \hspace{0.3cm} \textrm{.}
              \end{equation}
\par
In the inflation phase the function $f_{k}(k|\eta|)$ is defined as
              \begin{equation}
                 f_{k}(k|\eta|)=\frac{\sqrt{\pi}}{2}\sqrt{|\eta|}\mathcal{H}_{\nu}^{(2)}(k|\eta|)\quad,
              \end{equation}
and by using the small argument limit of the Hankel functions \cite{abramowitz}, we have
              \begin{equation}
                 P_{T}(k,\eta)=\frac{8\pi2^{2\nu}\Gamma^{2}(\nu)}{\pi^3}\frac{k^2}{a^2\phi}(k|\eta|)^{1-2v}
                 \hspace{0.3cm} \textrm{.}
              \end{equation}
         \par 
By definition the Hubble parameter is given by $H=\frac{\dot{a}}{a}=\frac{1}{a^2}\frac{\mathrm{d}a}{\mathrm{d}\eta}$. So, the conformal time in this model can be written as
              \begin{equation}
                 \label{eq:etaBD}
                 |\eta|=\left(\frac{2\omega+1}{2\omega-1} \right)\frac{1}{aH}
                 \hspace{0.2cm} \textrm{.}
              \end{equation}
Thus, the power spectrum in the BDT during inflation for the modes out of the horizon becomes
\begin{equation}
                 P_{T}(k,\eta)=
                \frac{8\pi2^{2\nu}\Gamma^{2}(\nu)}{\pi^3}\left(\frac{2\omega+1}{2\omega-1} \right)^{1-2\nu}
                 \frac{H^2}{\phi}\left(\frac{k}{aH}\right)^{3-2\nu}
                 \hspace{0.2cm} \textrm{.}
\end{equation}

\par
In Figures~\ref{espec01}-\ref{espec05} we show the behavior of the power spectrum $P_T(k, \eta)$ as a function of the conformal time $|\eta|$, the frequency $f(\mbox{Hz})$ and of the parameter $\omega$.

\subsubsection{Spectral Index}
      
         \par
The spectral index is obtained by the expression
              \begin{equation}
                 n_{T}=\frac{\mathrm{d}{\ln{P_{T}(k,\eta)}}}{\mathrm{d}\ln{k}}\Bigg|_{\alpha} 
                 \hspace{0.2cm} \textrm{,}
              \end{equation}     
where $\alpha\equiv aH=\left(\frac{2\omega+1}{2\omega-1}\right)k$ is the moment when the tensorial modes cross the horizon again. For this calculation we introduce the parameter $\epsilon$, that measures the rate of decrease of the Hubble parameter during the inflation phase and it is given by
               \begin{equation}
                  \label{eq:epsilon}
                  \epsilon\equiv \frac{\mathrm{d}}{\mathrm{d}t}\left(\frac{1}{H}\right)=
                       -\frac{1}{aH^2}\frac{\mathrm{d}H}{\mathrm{d}\eta}
                       \hspace{0.2cm} \textrm{.}
               \end{equation}
          \par    
In the JBDF we obtain that
               \begin{equation}
                  \label{eq:epsilonBD}
                  \epsilon=\frac{2}{2\omega+1}
                  \hspace{0.2cm} \textrm{.}
               \end{equation}  
         \par   
Thus, we have
\begin{equation}
\label{espectral}
                  n_{T}=\frac{\mathrm{d}}{\mathrm{d}\ln{k}}\left[\ln{\left(\frac{H^2}{\phi}\right)}\right]
                  \Bigg|_{\alpha}=
                 \biggl(2\frac{k}{H}\frac{\mathrm{d}H}{\mathrm{d}k}
                  -\frac{\mathrm{d}\ln{\phi}}{\mathrm{d}\ln{k}}\biggr)\Bigg|_{\alpha}
                  \hspace{0.2cm} \textrm{.}
\end{equation}
         \par
The first term on the right side is calculated as follows
\begin{equation}
                   \label{eq:Hetak}
                   2\frac{k}{H}\frac{\mathrm{d}H}{\mathrm{d}k}\Bigg|_{\alpha}=
                      2\frac{k}{H}\frac{\mathrm{d}H}{\mathrm{d}|\eta|}\frac{\mathrm{d}|\eta|}{\mathrm{d}k}
                      \Bigg|_{\alpha}=
                     2\frac{k}{H}\frac{\mathrm{d}H}{\mathrm{d}\eta}
                      \left(\frac{\mathrm{d}|\eta|}{\mathrm{d}\eta}\right)^{-1}\frac{\mathrm{d}|\eta|}{\mathrm{d}k}
                      \Bigg|_{\alpha}
                      \hspace{0.2cm} \textrm{,}
\end{equation}
and, by using the equation (\ref{eq:etaBD}), we have
\begin{equation}
                  \label{eq:etak}
                  \frac{\mathrm{d}|\eta|}{\mathrm{d}k}\Bigg|_{\alpha}=
                     \frac{\mathrm{d}}{\mathrm{d}k}\left(\left(\frac{2\omega+1}{2\omega-1}\right)\frac{1}{aH}\right)
                     \Bigg|_{\alpha}=
                     \frac{\mathrm{d}}{\mathrm{d}k}\left(\frac{1}{k}\right)=-\frac{1}{k^{2}}
                     \hspace{0.2cm} \textrm{.}
\end{equation}
\par 
In this way, with the help of the equations (\ref{eq:epsilonBD}), (\ref{eq:Hetak}) and (\ref{eq:etak}), the equation (\ref{espectral}) can be written as
               \begin{equation}
                  \label{eq:indiceomegaphi}
                  n_{T}=-2\epsilon\left(\frac{2\omega+1}{2\omega-1}\right)-\frac{\mathrm{d}\ln{\phi}}{\mathrm{d}\ln{k}}
                  \Bigg|_{\alpha}
                  \hspace{0.2cm} \textrm{.}
               \end{equation}
\par
When $\omega\rightarrow\infty$ and considering $\phi$ as a constant we obtain the same result that in GR, \textit{i. e.}, $n_T = - 2\epsilon$  \cite{dodelson}.
         \par  
Likewise, we can calculate the term dependent on the scalar field $\phi$ in equation (\ref{eq:indiceomegaphi})
\begin{equation}
                  \frac{\mathrm{d}\ln{\phi}}{\mathrm{d}\ln{k}}\Bigg|_{\alpha}=
                  \frac{k}{\phi}\frac{\mathrm{d}\phi}{\mathrm{d}k}\Bigg|_{\alpha}=
                  \frac{k}{\phi}\frac{\mathrm{d}\phi}{\mathrm{d}\eta}\frac{\mathrm{d}\eta}{\mathrm{d}k}
                  \Bigg|_{\alpha}=
                  \frac{k}{\phi}\frac{\mathrm{d}\phi}{\mathrm{d}\eta}\left(\frac{\mathrm{d}|\eta|}{\mathrm{d}\eta}\right)^{-1}
                  \frac{\mathrm{d}|\eta|}{\mathrm{d}k}\Bigg|_{\alpha}  
                  \hspace{0.2cm} \textrm{,}
\end{equation}
that with the help of equation (\ref{eq:etaBD}) and with the background solutions of the scalar field we obtain the following result for the spectral index $n_T$
               \begin{equation}
                  n_{T}=-2\epsilon\left(\frac{2\omega+1}{2\omega-1}\right)-\frac{4}{2\omega-1}
                  \hspace{0.2cm} \textrm{,}
               \end{equation}   
Again, the above result gives the same value of the spectral index of the tensorial modes of GR at the time of inflation, when $\omega
\rightarrow \infty $.
\par
We can write the expression of $n_T$ in terms of the parameter $\omega$. In order to do that we replace the relation (\ref{eq:epsilonBD}) in the result found above, such that
               \begin{equation}
                  \label{eq:indiceomega}
                  n_{T}=\frac{-8}{2\omega-1} \hspace{0.2cm} \textrm{.}
               \end{equation}
                 
         \par
By using (\ref{eq:epsilon}) we obtain
               \begin{equation}
                  2\omega=\frac{2}{\epsilon}-1
                  \hspace{0.2cm} \textrm{,}
               \end{equation}      
which can be replaced in the final result for the spectral index such that
               \begin{equation}
                  n_{T}=\frac{4\epsilon}{\epsilon-1}=-4\epsilon(1-\epsilon)^{-1}
                  \hspace{0.2cm} \textrm{.}
               \end{equation}
For small values of $\epsilon$ we can expand the term $(1-\epsilon)^{-1}$ obtaining
               \begin{equation}
                  n_{T}= -4\epsilon(1+\epsilon+\mathcal{O}(\epsilon^2))
                  \hspace{0.2cm} \textrm{,}
               \end{equation}
and considering only zero-order terms we have
               \begin{equation}
                  n_{T}\approx-4\epsilon
                  \hspace{0.2cm} \textrm{.}
               \end{equation}

\subsubsection{Energy Density}
      
         \par
From the Lagrangian density (\ref{lagran01}) we obtain the energy-momentum tensor of the gravitational waves
               \begin{equation}
                  {T}_{\mu\nu}=-\frac{\phi}{8\pi}\frac{1}{4}
                  \left(\partial_{\mu}h_{ij}\partial_{\nu}h^{ij}-
                  \frac{1}{2}g_{\mu\nu}g^{\alpha\beta}\partial_{\alpha}h_{ij}\partial_{\beta}h^{ij}\right)
                  \hspace{0.2cm} \textrm{.}
               \end{equation}  
         \par
Considering 
               \begin{equation}
                  h_{ij}=\sqrt{16\pi}\sum_{\lambda = \otimes , \oplus}{\epsilon_{ij}{^{(\lambda)}}h^{(\lambda)}}
                  \hspace{0.3cm} \textrm{,} \hspace{0.2cm}
                  h^{(\lambda)}=\frac{\mu^{(\lambda)}}{a\sqrt{\phi}}
                  \hspace{0.2cm} \textrm{,}
               \end{equation}
we find that
               \begin{equation}
                  {T}_{\mu\nu}=-2\phi\left(\partial_{\mu}h\partial_{\nu}h-
                     \frac{1}{2}g_{\mu\nu}g^{\alpha\beta}\partial_{\alpha}h\partial_{\beta}h\right)
                     \hspace{0.2cm} \textrm{,}
               \end{equation}
where
               \begin{equation}
                  h=h^{\otimes}=h^{\oplus}
                  \hspace{0.2cm} \textrm{.}
               \end{equation}
         \par
The energy density is the time component of the energy-momentum tensor of the gravitational waves
               \begin{equation}
                  \rho={T}_{0}^{~0}=\phi\left(\frac{h'^{2}}{a^2}+
                       g^{ij}\partial_{i}h\partial_{j}h\right)
                  \hspace{0.2cm} \textrm{.}
               \end{equation}
         \par
To calculate the quantum energy density we first consider the following redefinition
               \begin{equation}
                  h\rightarrow\hat{h}=\frac{\hat{\mu}}{a\sqrt{\phi}}
                  \hspace{0.2cm} \textrm{,}
               \end{equation}
such that the expectation value of the energy density is given by
               \begin{equation}
                  \left\langle\rho\right\rangle=\left\langle0\right|\rho\left|0\right\rangle=
                  \frac{\phi}{a^{2}}\left(\left\langle0\right|\hat{h}'\hat{h}'^{*}\left|0\right\rangle+
                  \left\langle0\right|\nabla\hat{h}\cdot\nabla\hat{h}^{*}\left|0\right\rangle\right)
                  \hspace{0.2cm} \textrm{.}
               \end{equation}
         \par
With help of the expansions
\begin{eqnarray}
                 \label{eq:operadores}
                 \hat{\mu}=\frac{1}{2}\int{\frac{\mathrm{d}^{3}k}{(2\pi)^{3/2}}\left[\hat{\mu}_{\vec{k}}(\eta)e^{-\dot{\imath}\vec{k}\cdot\vec{x}} 
                 +\hat{\mu}_{\vec{k}}^{\dagger}(\eta)e^{\dot{\imath}\vec{k}\cdot\vec{x}} \right]} \quad, \nonumber \\
                 \hat{\pi}=\frac{1}{2}\int{\frac{\mathrm{d}^{3}k}{(2\pi)^{3/2}}\left[\hat{\pi}_{\vec{k}}(\eta)e^{-\dot{\imath}\vec{k}\cdot\vec{y}} 
                 +\hat{\pi}_{\vec{k}}^{\dagger}(\eta)e^{\dot{\imath}\vec{k}\cdot\vec{y}} \right]} \hspace{0.2cm} \textrm{,}
\end{eqnarray}
the solutions
\begin{eqnarray} 
                 \label{eq:solgeloperadores}
                 \hat{\mu}_{\vec{k}}=\hat{a}_{\vec{k}}(\eta_0)f_{k}(\eta)+\hat{a}_{-\vec{k}}^{\dagger}(\eta_0)f_{k}^{*}(\eta) \quad, \nonumber \\
                 \hat{\pi}_{\vec{k}}=\hat{a}_{\vec{k}}(\eta_0)g_{k}(\eta)+\hat{a}_{-\vec{k}}^{\dagger}(\eta_0)g_{k}^{*}(\eta)
\hspace{0.2cm} \textrm{,}
\end{eqnarray}
and the commutation relations
\begin{equation}
                 \label{eq:comutacao1}
                 \left[\hat{a}_{\vec{k}},\hat{a}_{\vec{q}}^{\dagger} \right]=\delta^{(3)}(\vec{k}-\vec{q})   \hspace{0.2cm} \textrm{,}
\end{equation}
we obtain the expectation value of the energy density of the gravitational waves in the BDT that can be write as
\begin{equation}
                 \left\langle\rho\right\rangle=
                 \frac{1}{a^4}\int \frac{k^{2}\mathrm{d}k}{(2\pi)^{3}}\biggl[
                       |g_{k}|^{2}+\left(k^{2}+\left(\frac{\Phi'}{\Phi}\right)^{2}\right)|f_{k}|^{2}-
                    \frac{\Phi'}{\Phi}\left(g_{k}f_{k}^{*}+g_{k}^{*}f_{k}\right)\biggr]\frac{2\sin{kr}}{kr}
                       \hspace{0.2cm} \textrm{,}
\end{equation} 
where we have considered the presence of an isotropic background of relic graviton and $r=|\vec{x}-\vec{y}|$. In the limit $\vec{x}\rightarrow\vec{y}$ we have
\begin{equation}
                  \left\langle\rho\right\rangle=
                  \frac{1}{a^4}\int \frac{k^{2}\mathrm{d}k}{2\pi^{2}}\biggl[
                       |g_{k}|^{2}+\left(k^{2}+\left(\frac{\Phi'}{\Phi}\right)^{2}\right)|f_{k}|^{2}-
                       \frac{\Phi'}{\Phi}\left(g_{k}f_{k}^{*}+g_{k}^{*}f_{k}\right)\biggr]
                       \hspace{0.2cm} \textrm{.}
\end{equation}
\par
The spectral energy density per logarithmic interval of the wavenumber  is defined as
               \begin{equation}
                  \Omega(k,\eta)=\frac{1}{\rho_{c}}\frac{\mathrm{d}\left\langle\rho\right\rangle}{\mathrm{d}\ln{k}}
                  \hspace{0.2cm} \textrm{,}
               \end{equation}
\textit{i. e.},
\begin{equation}
                  \Omega(k,\eta)=\frac{8\pi\phi^{-1}}{3H_{0}^{2}}\frac{k^{3}}{2\pi^{2}a^{4}}
                  \biggl[|g_{k}|^{2}+ \left(k^{2}+\left(\frac{\Phi'}{\Phi}\right)^{2}\right)|f_{k}|^{2}-
                  \frac{\Phi'}{\Phi}\left(g_{k}f_{k}^{*}+f_{k}g_{k}^{*}\right)\biggr]
                  \hspace{0.2cm} \textrm{,}
\end{equation}
where $\rho_c$ is the critical density of the Universe.                         
\par
For all the oscillation modes that are inside the Hubble radius (\textit{i. e.} $k\eta >> 1$), we have $g_ {k}\approx\mp\dot\imath kf_ {k}$. This implies that
               \begin{equation}
                  \Omega(k,\eta)\approx\frac{8\pi k^5}{3\pi^2H_{0}^2\phi a^4}|f_{k}|^2
                  \left(1+\frac{1}{2k^2}\left(\frac{\Phi'}{\Phi}\right)^2\right)
                  \hspace{0.2cm} \textrm{,}
               \end{equation}
and, rewriting it in terms of the power spectrum we have
               \begin{equation}
                  \Omega(k,\eta)\approx\frac{k^2}{12H_{0}^2a^2}P_{T}\left(1+\frac{1}{2k^2}\left(\frac{\Phi'}{\Phi}\right)^2\right)
                  \hspace{0.2cm} \textrm{.}
               \end{equation}
          \par
For modes outside the Hubble radius (\textit{i. e.} $k\eta<<1$) we have $g_{k}\approx-\frac{\Phi'}{\Phi}f_{k}$. This result allow us to find that
               \begin{equation}
                  \Omega(k,\eta)\approx \frac{8\pi k^5}{6\pi^2 H_{0}^2\phi a^4}|f_{k}|^2
                  \hspace{0.2cm} \textrm{,}
               \end{equation}
which can also be rewritten as a function of the power spectrum $P_T$
               \begin{equation}
                  \Omega(k,\eta)\approx\frac{k^2}{24H_{0}^2a^2}P_{T}(k,\eta)
                  \hspace{0.2cm} \textrm{.}
               \end{equation}
\par
\begin{figure}[h!]
  \centering
      \includegraphics[width=0.9\textwidth]{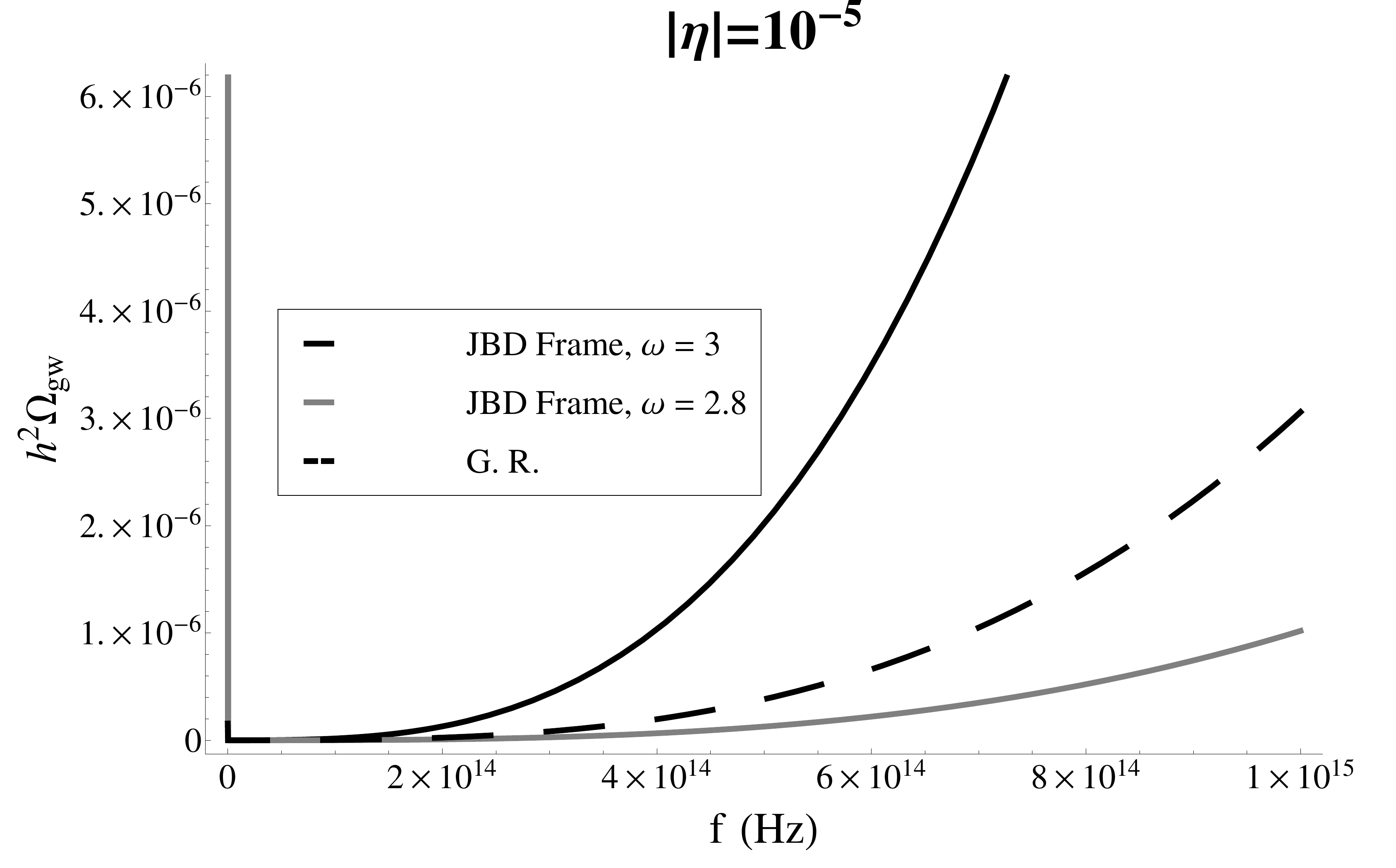}
  \caption{The graviton energy density in the BDT in the JBDF and in GR as a function of the frequency $f (\mbox{Hz})$, for small values of  $\omega$ and fixed $|\eta| = 10^{-5}$.}
\label{ener01}
\end{figure}

We stress that when the modes re-enter the horizon the energy density is proportional to the power spectrum. The behavior of the spectral energy density in the JBDF is shown in Figure~\ref{ener01}.

\section{Einstein frame}
\label{einstein}
In this section we calculate de general behavior of the BDT but now in the EF. As in all scalar-tensor theories, the action of the BDT is preserved under a group of field redefinitions that contains two functional degrees of freedom. Specifically, if one defines a new metric $\bar g_{\mu\nu}$ and a new scalar field $\bar\phi$ by     
      \begin{eqnarray}
\label{transf01}
         \bar{g}_{\mu\nu} &=& \Omega^{2}g_{\mu\nu}\nonumber\quad,\\
         \bar\phi &=& \frac{e^{A\sigma}}{2\kappa^2}\quad, 
      \end{eqnarray}
where $\Omega^{2}= e^{A\sigma}$ is a function of the coordinate and $\sigma$ is a new scalar field. We also consider the following definitions:
      \begin{equation}
         \kappa^{2}=8\pi G\hspace{0.3cm} \textrm{,} \hspace{0.3cm} A=\beta \kappa \hspace{0.3cm} \textrm{,} \hspace{0.3cm}
         \beta^{2}=\frac{2}{2\omega+3}\hspace{0.3cm} \textrm{.}
      \end{equation}
\subsection{Background solutions}
\label{subsectionbackE}

The conformal transformation (\ref{transf01}) change the theory's frame from the JBDF to the EF and the action can be written as
      \begin{equation}
         \label{eq:actionEF}
\bar{\mathcal{S}}=\int \mathrm{d}^{4}x{\sqrt{-\bar{g}}}\biggl[\frac{\bar{R}}{2\kappa^{2}}-\frac{1}{2}(\bar{g}^{\mu\nu}\sigma_{,\mu}\sigma_{,\nu})+\mathcal{L}_{mat}\biggr] \hspace{0.3cm}\textrm{,}
      \end{equation}
where this Lagrangian was obtained with the help of an integration that is necessary to eliminate a boundary contribution \cite{dicke01}. To obtain the
      fields equations we have to variate the action (\ref{eq:actionEF}) with respect to the metric tensor $\bar{g}^{\mu\nu}$ and the
      scalar field $\sigma$, and we get
      \begin{eqnarray}
         \label{eqn:fieldeqs}
         \bar{R}_{\mu\nu}-\frac{1}{2}\bar{g}_{\mu\nu}\bar{R}=\kappa^{2}\bar{T}_{\mu\nu}+\kappa^{2}\left(\sigma_{,\mu}\sigma_{,\nu}-
         \frac{1}{2}\bar{g}_{\mu\nu}\bar{g}^{\alpha\beta}\sigma_{,\alpha}\sigma_{,\beta}\right)\quad, \\
         \square \sigma = \frac{1}{2}Ae^{-2A\sigma}T\hspace{0.3cm} \textrm{,}
      \end{eqnarray} 
      where $\bar{T}_{\mu\nu}$ is the stress-energy tensor in the EF. It is related with the stress-energy tensor in the JBDF $T_{\mu\nu}$ by the expression $\bar{T}_{\mu\nu}=e^{-2A\sigma}T_{\mu\nu}$.
      \par
         The stress-energy tensor conservation gives us
         \begin{equation}
            \label{eq:conservation}
            \bar{T}^{\mu\nu}_{\hspace{10pt};\nu}+\bar{g}^{\mu\nu}\sigma_{\mu}\square \sigma=0 \hspace{0.3cm} \textrm{.}
         \end{equation}
         Applying the flat Friedmann-Robertson-Walker metric in the field equations (\ref{eqn:fieldeqs}) and in the conservation relation 
         (\ref{eq:conservation}) and considering a perfect fluid with equation of state $p=\gamma\rho$ getting
         \begin{eqnarray}
            \label{eq:field1}
            \frac{\dot{b}^2}{b^2}=\frac{\kappa^{2}}{2}\left(e^{-2A\sigma}\rho+\frac{\dot{\sigma}^2}{2}\right)\quad, \\
            \label{eq:field2}
            \ddot{\sigma}+3\frac{\dot{b}}{b}\dot{\sigma}=\frac{1}{2}Ae^{-2A\sigma}\left(3\gamma-1\right)\rho\quad, \\
            \label{eq:conservation1}
            \dot{\rho}+3\frac{\dot{b}}{b}(1+\gamma)\rho=\frac{3}{2}A\dot{\sigma}(1+\gamma)\rho\quad,
         \end{eqnarray}
where $b(t)$ is the scale factor in this scenario. The equation (\ref{eq:conservation1}) can be easily integrated to give us
         \begin{equation}
            \rho b^{3(1+\gamma)}e^{-\frac{3}{2}(1+\gamma)\sigma}=C_{2} \hspace{0.3cm} \textrm{,}
         \end{equation}
         where $C_{2}$ is an integration constant.
     \par
         In order to solve the equations (\ref{eq:field1}), (\ref{eq:field2}) and (\ref{eq:conservation1}) we will assume that the field $\sigma$ and
         the scale factor $b$ have the following power law solutions:
         \begin{equation}
            b(t)=b_0 t^{p} \hspace{0.3cm} \textrm{,} \hspace{0.3cm} e^{A\sigma}=e^{A\sigma_0} t^q \hspace{0.3cm} \textrm{.}
         \end{equation}
     \par
        For the case of inflation, when $\gamma=-1$, the solutions are
        \begin{equation}
           q=1 \hspace{0.1cm} \textrm{,} \hspace{0.3cm} p_{\pm}=-\frac{1}{4\beta^2}\left(1\pm \sqrt{1 + \frac{16\beta^2}{3}}\right)
           \hspace{0.1cm} \textrm{,} \hspace{0.3cm} \rho=C_2 \hspace{0.3cm} \textrm{.}
        \end{equation}
     \par
In the following we consider the $p_+$ solution because it reproduces the inflationary model in the limit $-\infty < \eta \leq -\eta_{1}$. Using the conformal time transformation $\mathrm{d}t=b(\eta)\mathrm{d}\eta$, where $t$ and $\eta$ are the cosmic and  the conformal times, respectively, we have
        \begin{equation}
           b(\eta)=b_{0}\left[b_{0}\left(p_{+}-1\right)(-\eta)\right]^{\frac{p_{+}}{1-p_{+}}} \hspace{0.1cm} \textrm{,} \hspace{0.3cm}
           -\infty < \eta \leq -\eta_{1} \hspace{0.3cm} \textrm{,}
        \end{equation} 
        where $-\eta_{1}$ is the time when the radiation era begins.

\subsection{The quantum gravitational waves equation}
\label{subsectionwavesE}
 We can perturb the metric tensor such that $\bar{g}_{\mu\nu}=\bar{\eta}_{\mu\nu}+\bar{h}_{\mu\nu}$, where $\bar{\eta}_{\mu\nu}$ is the
        background metric tensor and $\bar{h}_{\mu\nu}$ is the metric perturbation. Using the perturberd metric tensor in the field equation 
        (\ref{eqn:fieldeqs}) we obtain the gravitational waves equation in the EF:
        \begin{equation}
          \ddot{\bar{h}}_{ij}-\frac{\dot{b}}{b}\dot{\bar{h}}_{ij}+
             \left(-\frac{\bar{\nabla}}{b^{2}}-2\frac{\ddot{b}}{b}\right)\bar{h}_{ij}=0 \hspace{0.3cm} \textrm{,}
        \end{equation}
        where $\bar{\nabla}=\bar{\eta}^{ij}\partial_{i}\partial_{j}$.
     \par
        In order to solve the above equation it is convenient to make the transformation 
        \begin{equation}
           \bar{h}_{ij}(\vec{x},\eta)=\sqrt{2}
           \sum_{s=\oplus,\otimes}\kappa{\int{\frac{\mathrm{d}^{3}k}{(2\pi)^{3/2}}\epsilon_{ij}^{(s)}(\vec{k})\frac{\mu^{(s)}(\eta)}{b(\eta)}
           e^{-\dot{\imath}(\vec{k}\cdot\vec{x})}}} \hspace{0.3cm} \textrm{,}
        \end{equation}
        where $\vec{k}$ is the wave number and the polarization tensor $\epsilon_{ij}^{(s)}(\vec{k})$ obeys the relation
        $\epsilon_{ij}^{(s)}(\vec{k})\epsilon^{ij{}(s')}(\vec{k'})=2\delta(\vec{k}-\vec{k'})\delta^{(ss')}$. The wave equation is then
        \begin{equation}
           \label{eq:wave}
           \mu''+\left(k^{2}-\frac{b''}{b}\right)\mu=0 \hspace{0.3cm} \textrm{,}
        \end{equation}
        where $\mu'=\mathrm{d}\mu/\mathrm{d}\eta$ and
        \begin{equation}
           \frac{b''}{b}=\frac{F(\beta^2)}{\eta^2} \hspace{0.1cm} \textrm{,}
\end{equation}
 with
\begin{equation}
           F(\beta^2)=\left(\frac{p_{+}}{1-p_{+}}\right)\left(\frac{p_{+}}{1-p_{+}}-1\right)\textrm{.} 
\end{equation}
     \par
        To solve the wave equation in this frame we make the definitions $\mu\equiv\sqrt{\eta}g(z)$ where $z\equiv k\eta$ to obtain
        \begin{equation}
           z^{2}\frac{d^{2}g}{dz^{2}}+z\frac{dg}{dz}+\left(z^{2}-\left(F(\beta^2)-1/4\right)\right)g=0
           \hspace{0.3cm} \textrm{,}
        \end{equation}
        and the solution is
        \begin{equation}
           \mu(k\eta)=\sqrt{\eta}\left(A_{1}\mathcal{H}_{\nu}^{(1)}(k|\eta|)+A_{2}\mathcal{H}_{\nu}^{(2)}(k|\eta|)\right)
           \hspace{0.3cm} \textrm{,}
        \end{equation}
        for $\nu=\sqrt{F(\beta^2)-1/4}$.
     \par
        Following the quantization method described in \cite{mass} we transform the field $\mu$ into operator of the Heisenberg description of
        Quantum Mechanics:
        \begin{equation}
           \mu \rightarrow \hat{\mu}(\eta)=\frac{1}{2}\int{\frac{\mathrm{d}^{3}k}{(2\pi)^{3/2}}\left(\hat{\mu}_{\vec{k}}
             e^{-\dot{\imath}\vec{k}\cdot\vec{x}}+\hat{\mu}_{\vec{k}}^{\dagger}e^{\dot{\imath}\vec{k}\cdot\vec{x}}\right)}
             \hspace{0.3cm} \textrm{.}
        \end{equation}
        The coefficients $\hat{\mu}_{\vec{k}}$ can be written as
        \begin{equation}
           \hat{\mu}_{\vec{k}}=\hat{a}_{\vec{k}}(\eta_0)f_{k}(\eta)+\hat{a}_{-\vec{k}}^{\dagger}(\eta_0)f_{k}^{*}(\eta)
           \hspace{0.3cm} \textrm{,}
        \end{equation}
        where $\eta_0$ is a fixed comoving time and the functions $f_{k}(\eta)$ and the operators $\hat{a}_{\vec{k}}(\eta_0)$
        obeys the relations
        \begin{eqnarray}
           \label{eq:fkdif}
           f''_{k}+\left(k^2-\frac{b''}{b}\right)f_{k}=0 \hspace{0.3cm} \textrm{,} \\
           \label{eq:normafk}
           f'_{k}f^{*}_{k}-f_{k}f'^{*}_{k}=\dot{\imath} \hspace{0.3cm} \textrm{,} \\
           \label{eq:comutador}
           \left[\hat{a}_{\vec{k}},\hat{a}_{\vec{q}}^{\dagger}\right]=\delta^{(3)(\vec{k}-\vec{q})} \hspace{0.3cm} \textrm{.}
        \end{eqnarray}
        We have already found the solution for equations like equation (\ref{eq:fkdif}). To have a well behaved function in the limit 
        $\eta\rightarrow -\infty$ and with help of the normalization relation (\ref{eq:normafk}) we find that for inflation
        \begin{equation}
           f_{k}=\frac{\sqrt{\pi}}{2}e^{\dot{\imath}\theta_{k}}\sqrt{\eta}\mathcal{H}_{\nu}^{(2)}(k\eta)
           \hspace{0.3cm} \textrm{,}
        \end{equation} 
        where $\theta_{k}$ is a phase factor. Still following the prescription of \cite{mass} the number of gravitons is given by
        \begin{equation}
           N_k = \frac{|f_{k}|^2}{2k}\left(k^2+\left(\frac{b'}{b}\right)^2\right)+        
                 \frac{1}{2k}\left|\frac{\mathrm{d}f_{k}}{\mathrm{d}\eta}\right|^2-             
                 \frac{1}{2k}\frac{b'}{b}\left(f_{k}\frac{\mathrm{d}f^{*}_{k}}{\mathrm{d}\eta}+
                 f^{*}_{k}\frac{\mathrm{d}f_{k}}{\mathrm{d}\eta}\right) -\frac{1}{2}
                 \hspace{0.2cm} \textrm{,}
        \end{equation}
\par
In Figures~\ref{num01}-\ref{num02} we show the behavior of the number of particles in the inflation era in the BDT in the case of the EF to different values of $\omega$ and small and big values of the frequency and we compare with the JBDF and with the GR.

\subsection{Physical Observables}
\label{subsectionobsE}

\subsubsection{Power Spectrum}

 \par   
The quantization process gives us
        \begin{equation}
           \label{oper02}
                 \bar{h}_{ij}\rightarrow\hat{\bar{h}}_{ij}(\vec{x},\eta) =
                  \frac{\sqrt{2}\kappa}{b(\eta)}\sum_{\lambda=\otimes,\oplus}\frac{1}{2}
                    \int \frac{\epsilon_{ij}^{(\lambda)}(\vec{k})\mathrm{d}^3k}{(2\pi)^{3/2}}
                    \left(\hat{\mu}_{\vec{k},(\lambda)}
                    e^{-\dot{\imath}\vec{k}\cdot\vec{x}}
                    +\hat{\mu}_{\vec{k},(\lambda)}^{\dagger}e^{\dot{\imath}\vec{k}\cdot\vec{x}}\right)
                    \hspace{0.2cm} \textrm{,}
        \end{equation}
the same result obtained in GR. So, the gravitational waves power spectrum is given by
        \begin{equation}
           P_T=4\kappa^2\frac{k^3}{\pi^2}\frac{|f_k|^2}{b^2}
           \hspace{0.3cm} \textrm{.}
        \end{equation}
\par
\begin{figure}[h!]
  \centering
      \includegraphics[width=0.9\textwidth]{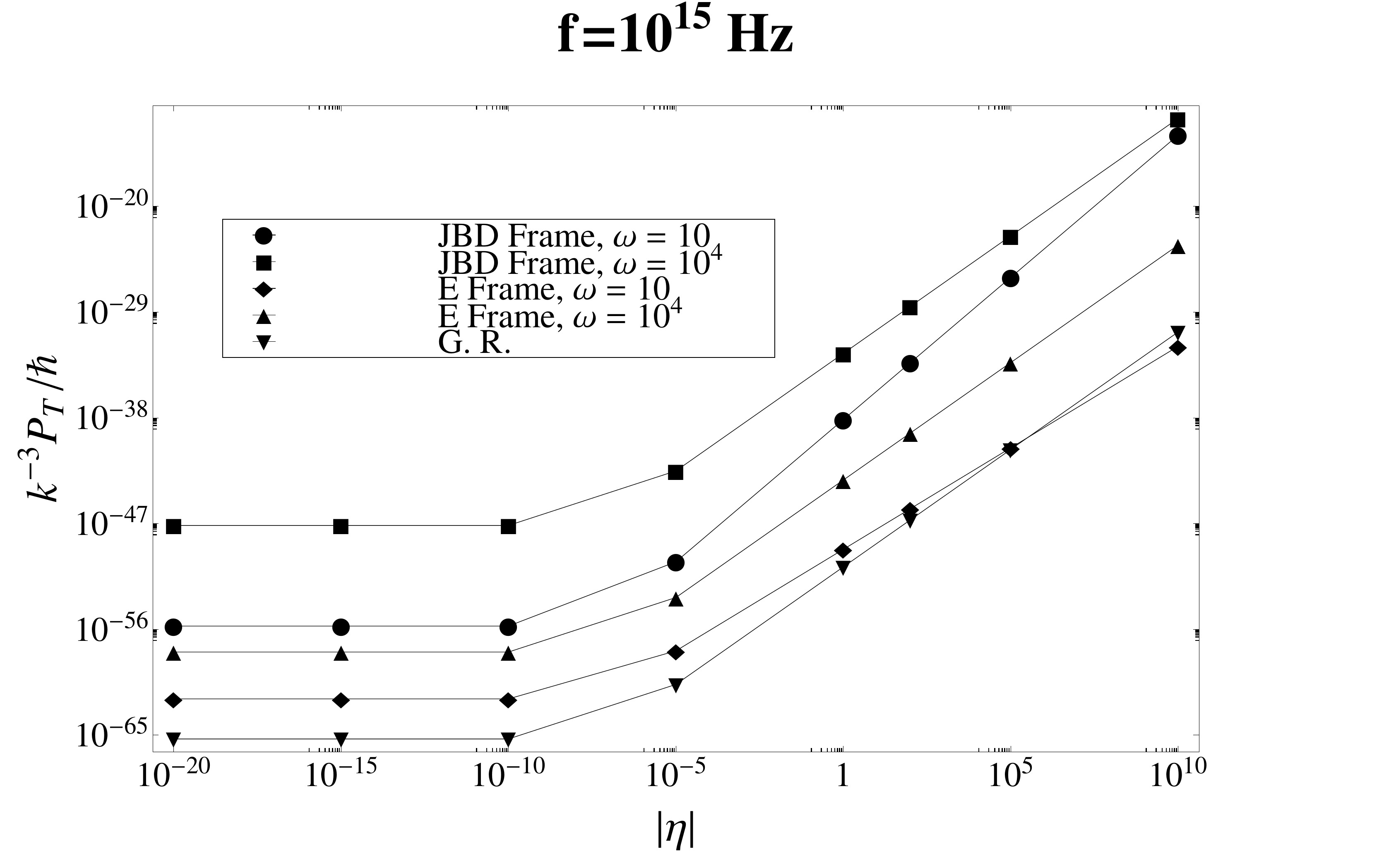}
  \caption{The power spectrum of the gravitational waves in the BDT and GR as a function of the conformal time $|\eta|$ with fix frequency $f = 10^{15}~\mbox{Hz}$ and large values of $\omega$ in both frames.}
\label{espec03}
\end{figure}
\begin{figure}[h!]
  \centering
      \includegraphics[width=0.9\textwidth]{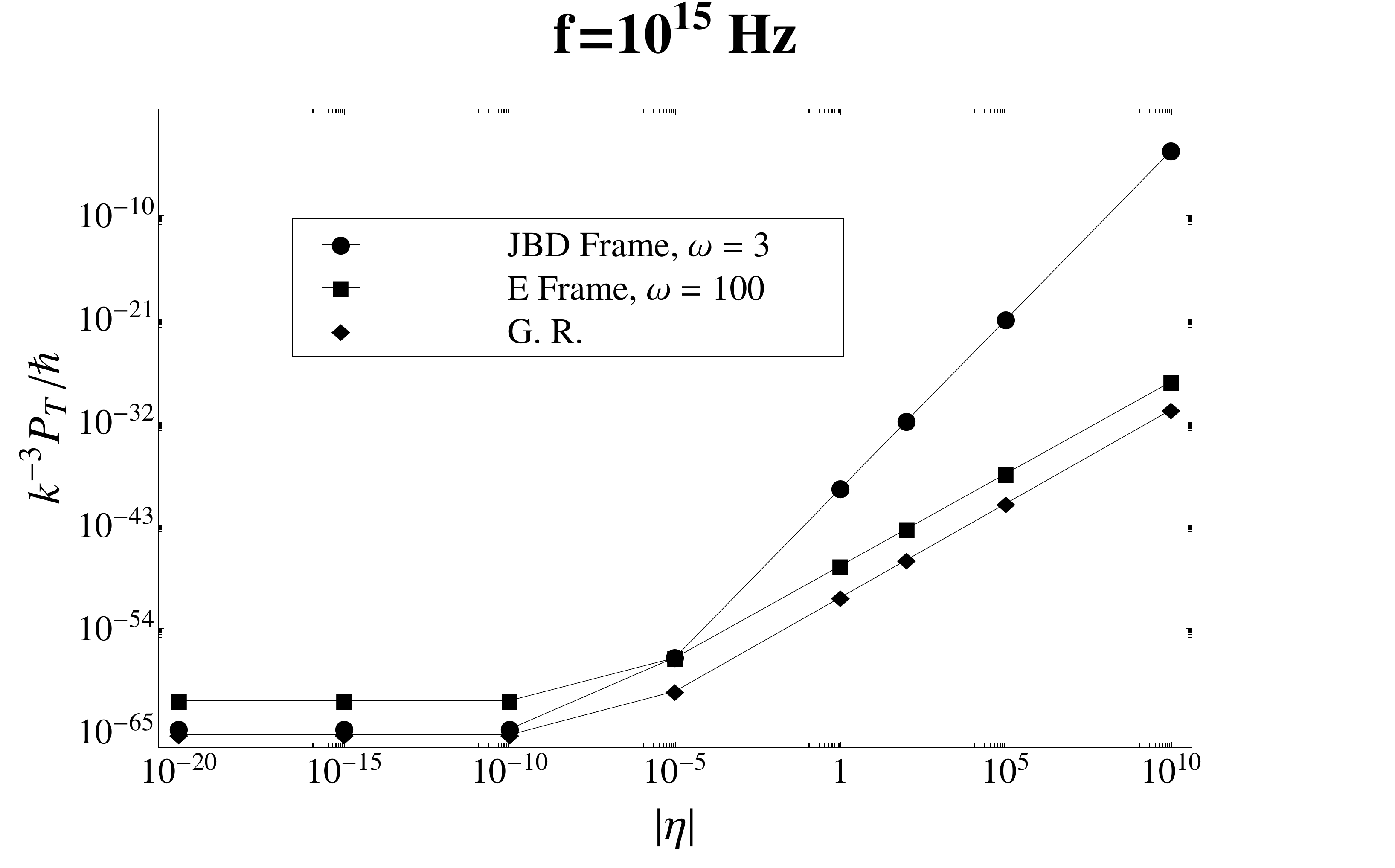}
  \caption{The power spectrum of the gravitational waves in the BDT and GR as a function of the conformal time $|\eta|$ with fix frequency $f = 10^{15}~\mbox{Hz}$ and some values of $\omega$ in both frames.}
\label{espec04}
\end{figure}
\begin{figure}[h!]
  \centering
      \includegraphics[width=0.9\textwidth]{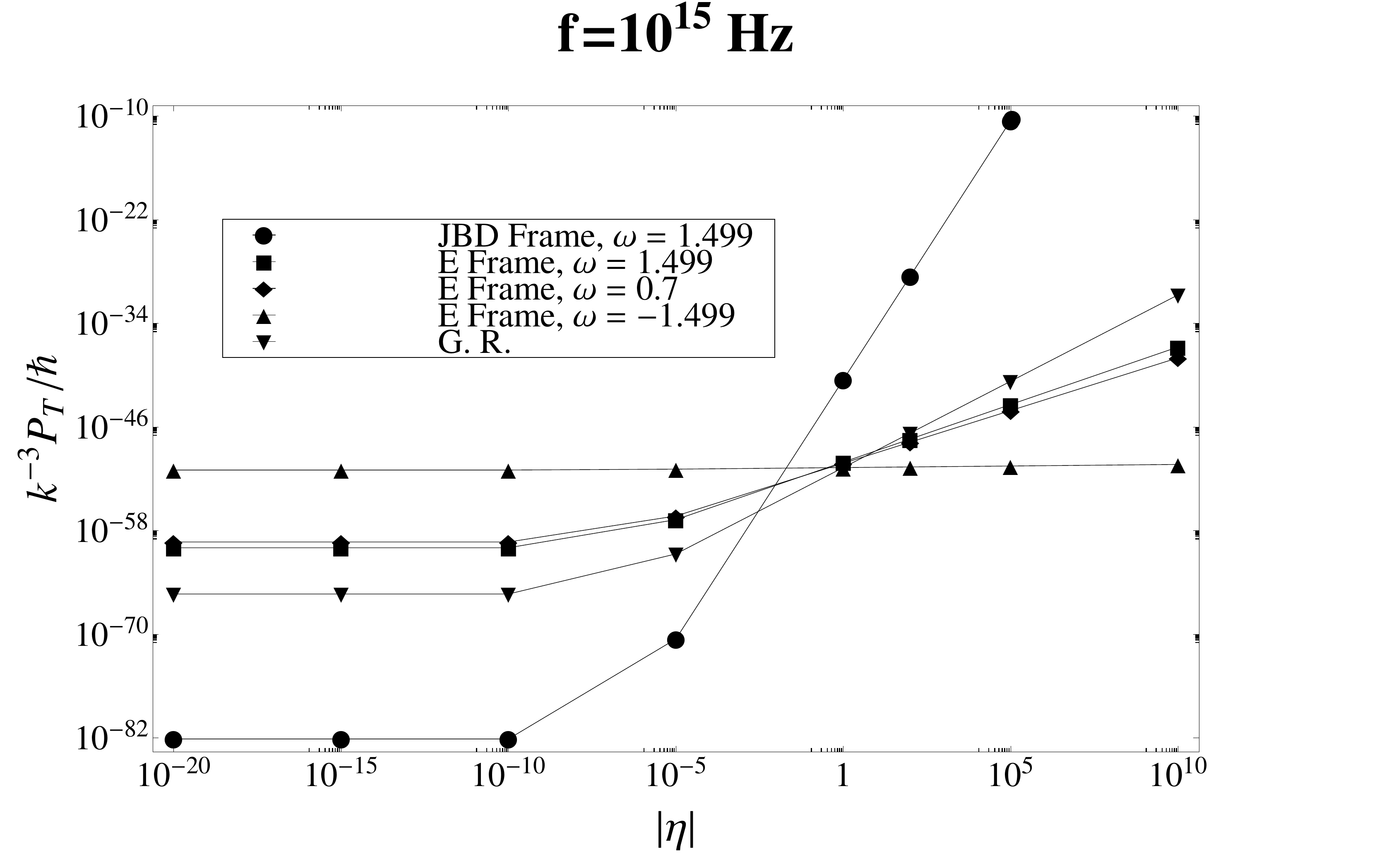}
  \caption{The power spectrum of the gravitational waves in the BDT and GR as a function of the conformal time $|\eta|$ with fix frequency $f = 10^{15}~\mbox{Hz}$ and some values of $\omega$ in both frames, including negative values in EF.}
\label{espec05}
\end{figure}

The behavior of the power spectrum is shown in Figures~\ref{espec01}-\ref{espec05}.

\subsubsection{Spectral Index}

\par
        In order to find the spectral index of the gravitational waves during inflation in the EF we have first to calculate the parameter $\epsilon$
        of inflation, given by
        \begin{equation}
           \epsilon = \frac{\mathrm{d}\bar{H}}{\mathrm{d}t}=-\frac{1}{b\bar{H}^2}\frac{\mathrm{d}\bar{H}}{\mathrm{d}\eta} = \frac{1}{p_{+}}
           \hspace{0.3cm} \textrm{,}
        \end{equation}  
        where $\bar{H} = \dot b/b$ is the Hubble parameter in the EF.
        During this calculation process, with help of the relation $b\bar{H}=\frac{\mathrm{d}b}{\mathrm{d}t}\frac{1}{b}$ it is possible to find
        \begin{equation}
           |\eta|=\left(\frac{p_{+}}{1-p_{+}}\right)\frac{1}{b\bar{H}}
           \hspace{0.3cm} \textrm{,}
        \end{equation} 
        that is necessary to calculate the spectral index. 
     \par
        The spectral index is given by
        \begin{equation}
           n_{T}=\frac{\mathrm{d}{\ln{P_{T}(k,\eta)}}}{\mathrm{d}\ln{k}}\Bigg|_{\alpha} =-2\epsilon \frac{p_+}{1-p_+} = -\frac{2}{1-p_+}
           \hspace{0.2cm} \textrm{,}
        \end{equation} 
        where $\alpha\equiv b\bar{H}=\left(\frac{p_{+}}{1-p_{+}}\right)k$ and the power spectrum $P_{T}$ is evaluated for $k\eta<<1$. We found that when $\omega\rightarrow\pm\infty$ the spectral index goes to $2\epsilon$ differently from the JBDF where, in the same limits, $n_T\rightarrow -2\epsilon$.

\subsubsection{Energy Density}

 \par
        The energy-momentum tensor of the gravitational waves is
               \begin{equation}
                  {T}_{\mu\nu}=-\frac{1}{\kappa^2}\frac{1}{4}
                  \left(\partial_{\mu}\bar{h}_{ij}\partial_{\nu}\bar{h}^{ij}-
                  \frac{1}{2}\bar{\eta}_{\mu\nu}\bar{\eta}^{\alpha\beta}\partial_{\alpha}\bar{h}_{ij}\partial_{\beta}\bar{h}^{ij}\right)
                  \hspace{0.2cm} \textrm{.}
               \end{equation} 
        Using the quantization (\ref{oper02}) we find
        \begin{equation}
                  \left\langle\rho\right\rangle= \left\langle0\right|\rho\left|0\right\rangle=
                  \frac{1}{b^4}\int \frac{k^{2}\mathrm{d}k}{2\pi^{2}}\biggl[
                       |g_{k}|^{2}+\left(k^{2}+\left(\frac{b'}{b}\right)^{2}\right)|f_{k}|^{2}-
                       \frac{b'}{b}\left(g_{k}f_{k}^{*}+g_{k}^{*}f_{k}\right)\biggr]
                       \hspace{0.2cm} \textrm{,}
        \end{equation}
        where $g_k=f'_k$ and $\left|0\right\rangle$ is the vacuum state. 
     \par
       The spectral energy density per logarithmic interval of the wavenumber  is defined as
               \begin{equation}
                  \Omega(k,\eta)=\frac{1}{\rho_{c}}\frac{\mathrm{d}\left\langle\rho\right\rangle}{\mathrm{d}\ln{k}}
                  \hspace{0.2cm} \textrm{,}
               \end{equation}
       \textit{i. e.},
       \begin{equation}
                  \Omega(k,\eta)=\frac{\kappa^2}{3\bar{H}_{0}^{2}}\frac{k^{3}}{2\pi^{2}b^{4}}
                  \biggl[|g_{k}|^{2}+ \left(k^{2}+\left(\frac{b'}{b}\right)^{2}\right)|f_{k}|^{2}-
                  \frac{b'}{b}\left(g_{k}f_{k}^{*}+f_{k}g_{k}^{*}\right)\biggr]
                  \hspace{0.2cm} \textrm{,}
       \end{equation}
       where $\rho_c$ is the critical density of the Universe.
\par
\begin{figure}[h!]
  \centering
      \includegraphics[width=0.9\textwidth]{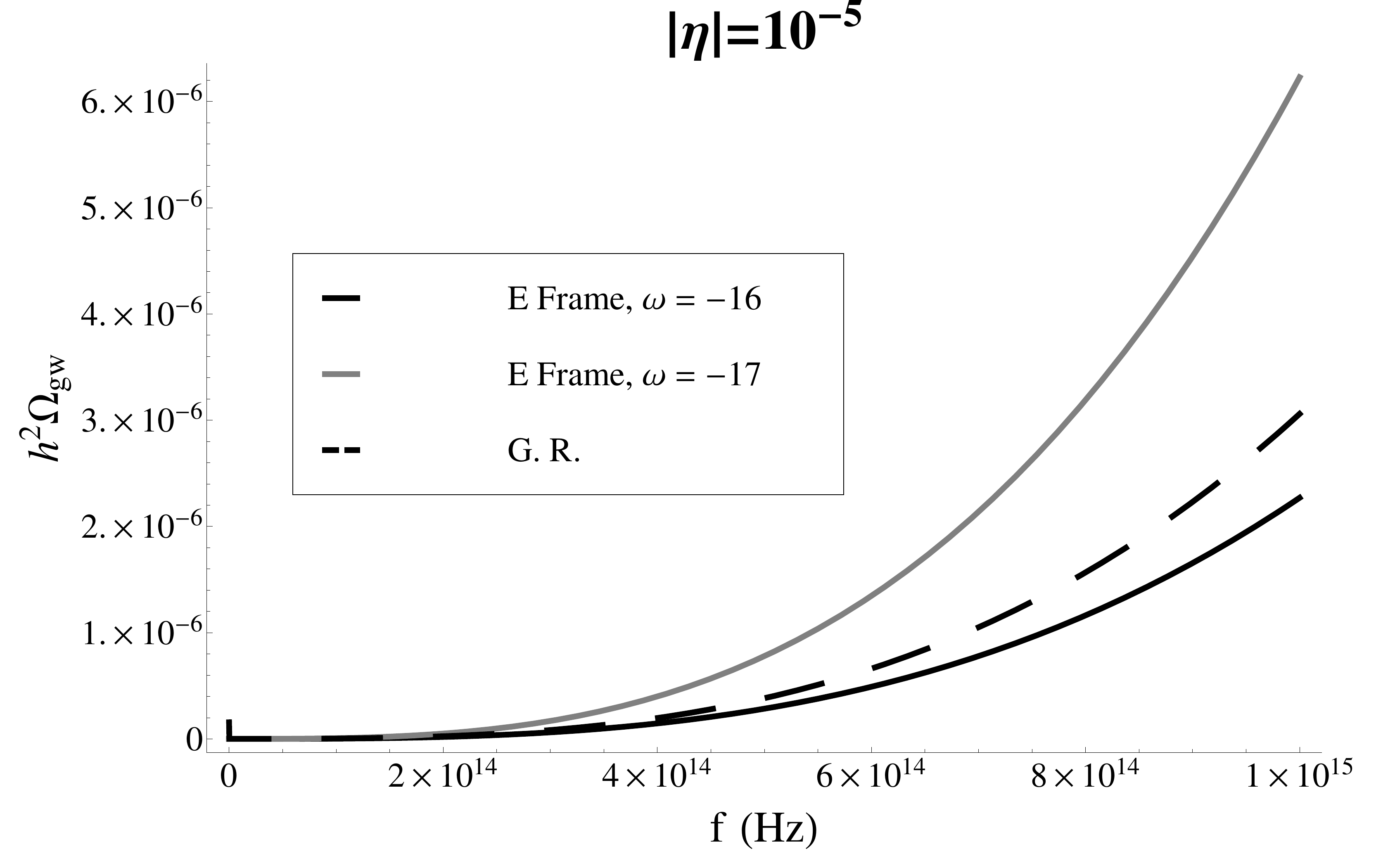}
  \caption{The graviton energy density in the BDT in the EF and in GR as a function of the frequency $f (\mbox{Hz})$, for negative values of  $\omega$ and fixed $|\eta| = 10^{-5}$.}
\label{ener02}
\end{figure}

The density energy behavior is shown in the EF in Figure~\ref{ener02}.

\section{Conclusions}
\label{conclusions}

\begin{figure}[h!]
  \centering
      \includegraphics[width=0.9\textwidth]{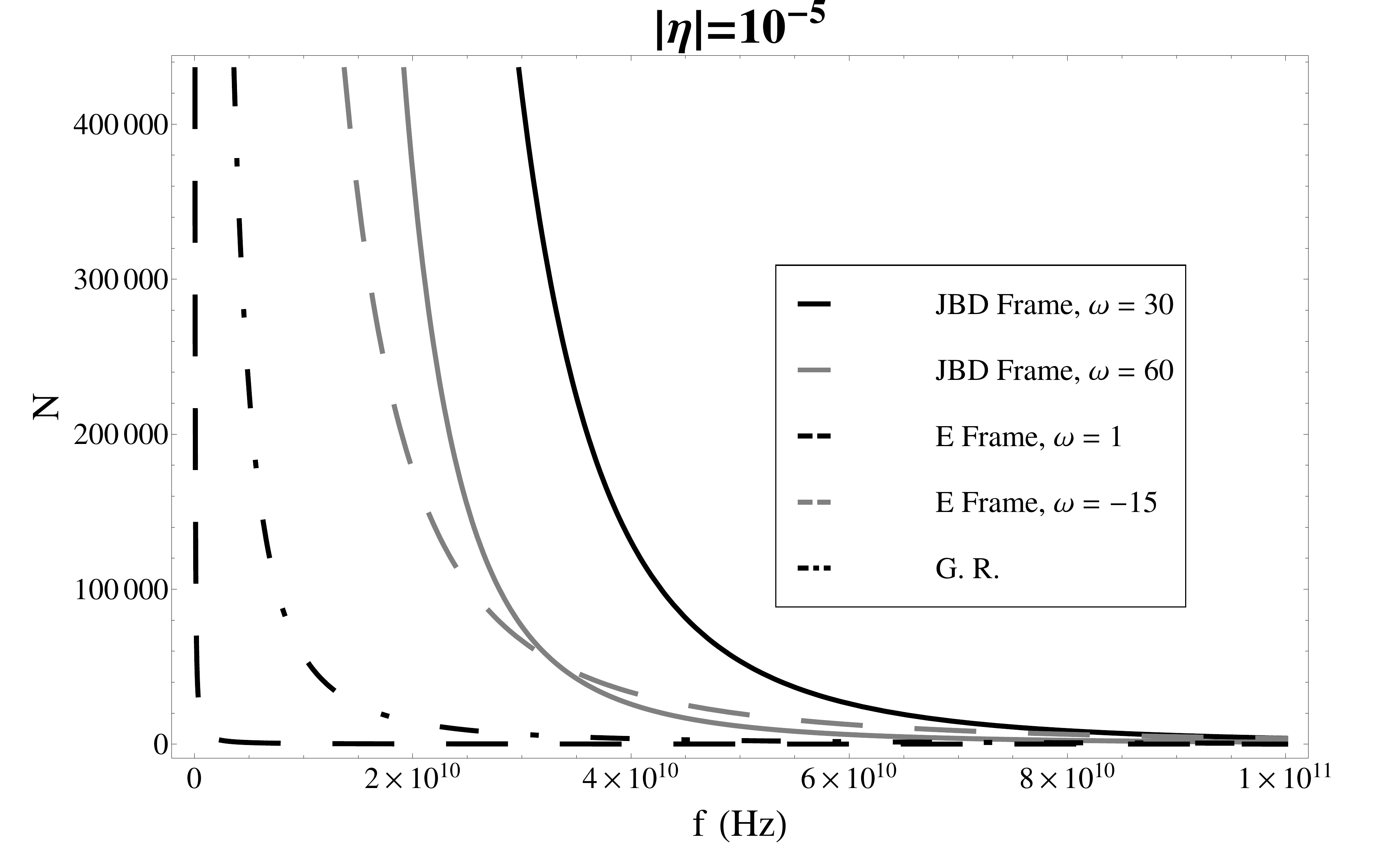}
  \caption{The number of gravitons produced by inflation in the BDT in the different frames and in the GR as a function of the frequency $f (\mbox{Hz})$ with fix conformal time $|\eta| = 10^{-5}$ for different values of $\omega$, including negative values in the EF.}
\label{num01}
\end{figure}

\begin{figure}[h!]
  \centering
      \includegraphics[width=0.9\textwidth]{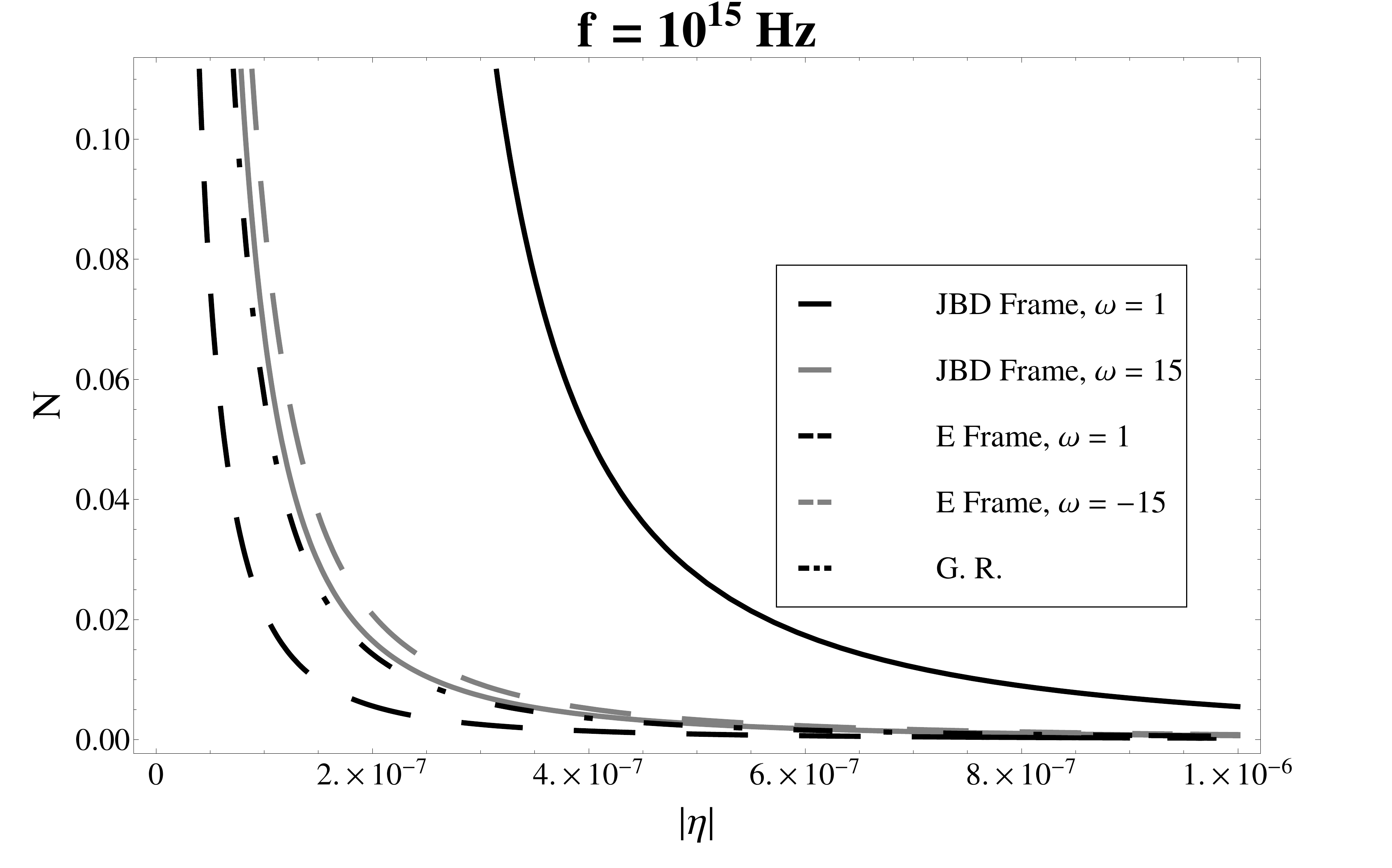}
  \caption{The number of gravitons produced by inflation in the BDT in the different frames and in the GR as a function of the conformal time $|\eta|$ with fix frequency $f = 10^{15}~\mbox{Hz}$ for different values of $\omega$, including negative values in the EF.}
\label{num02}
\end{figure}

\begin{figure}[h!]
  \centering
      \includegraphics[width=0.9\textwidth]{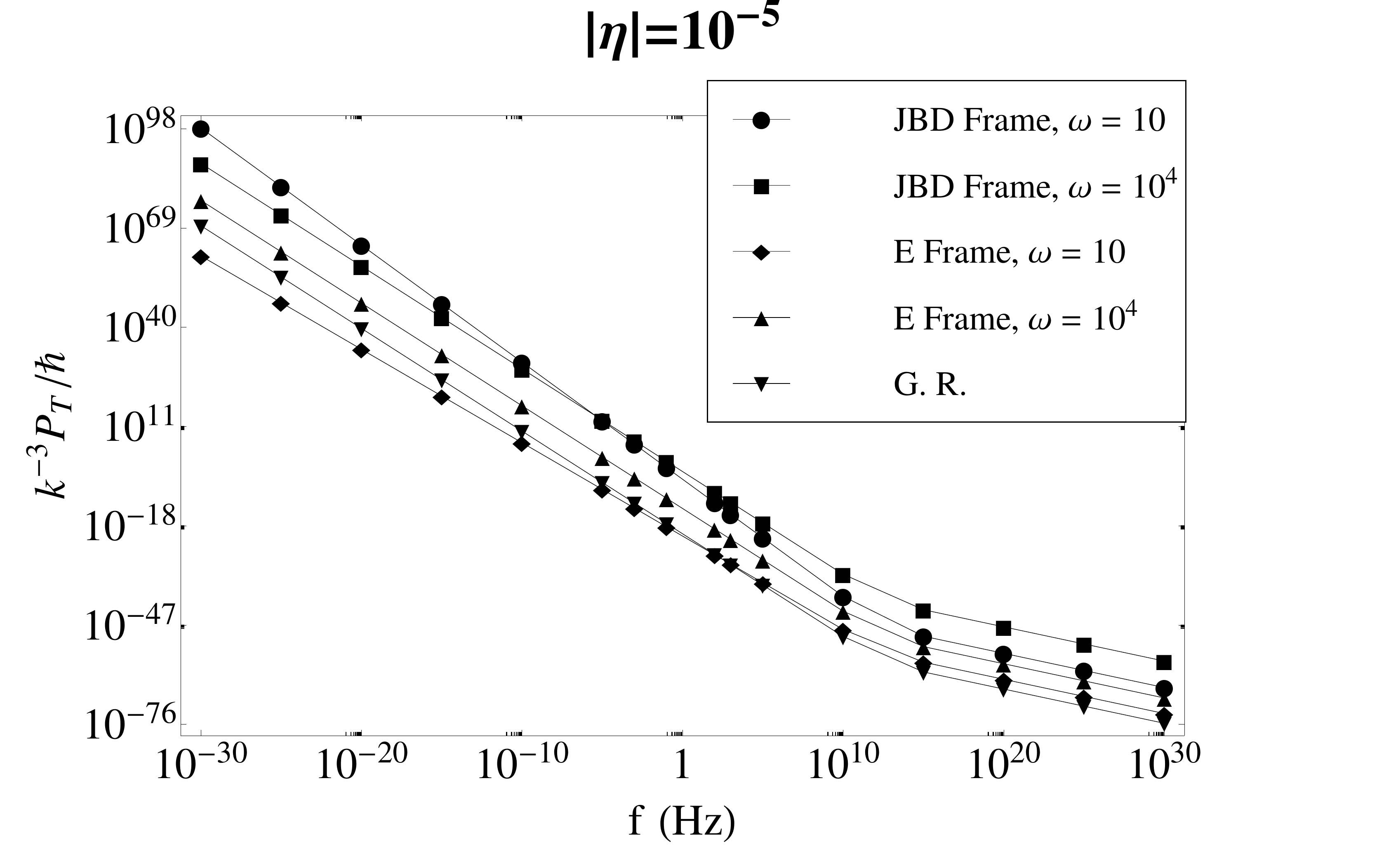}
  \caption{The power spectrum of the gravitational waves in the BDT and GR, in logarithmic scale, as a function of the frequency $f (\mbox{Hz})$ with fix conformal time $|\eta| = 10^{-5}$ and different values of $\omega$ in both frames.}
\label{espec01}
\end{figure}

\begin{figure}[h!]
  \centering
      \includegraphics[width=0.9\textwidth]{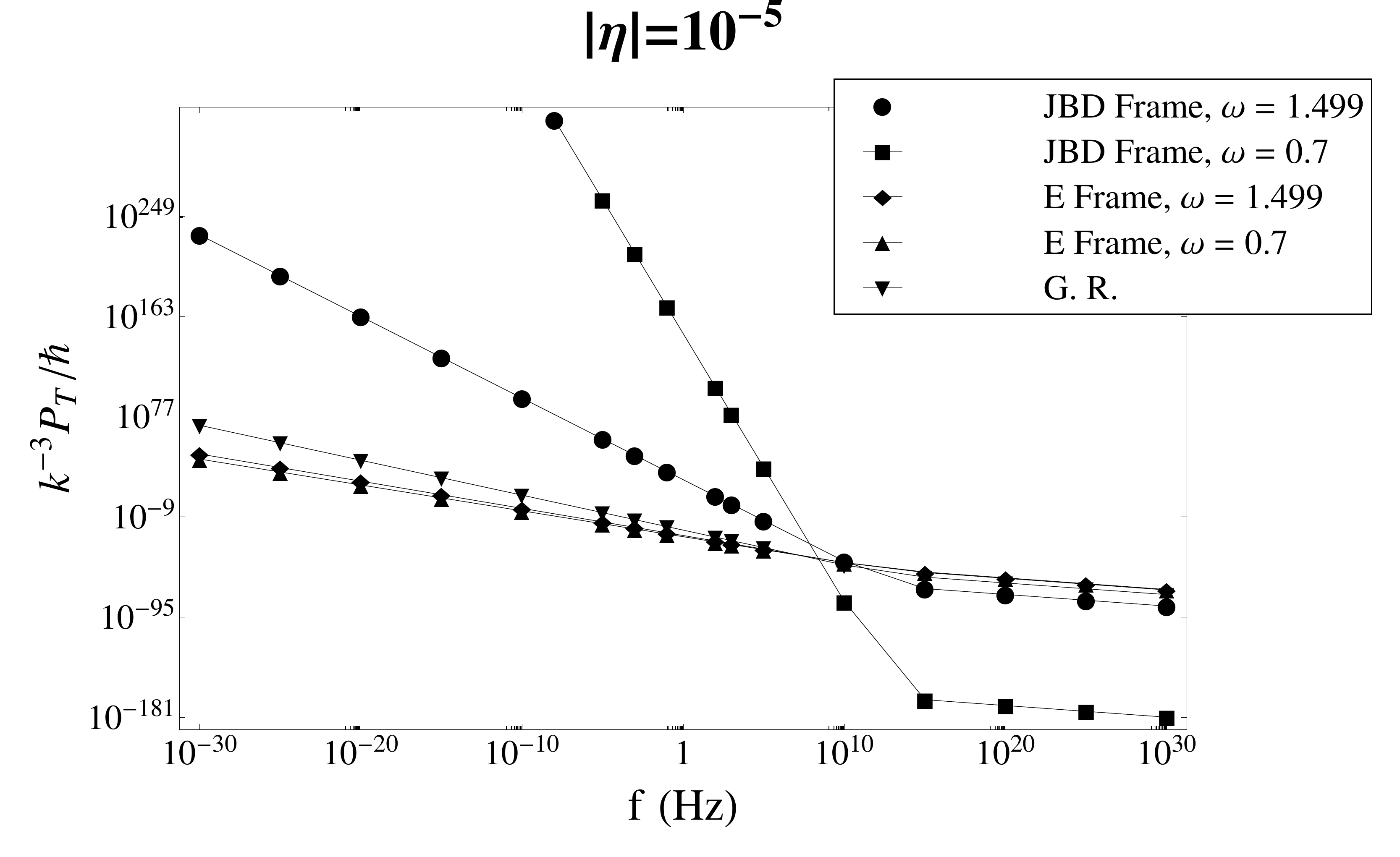}
  \caption{The power spectrum of the gravitational waves in the BDT and GR as a function of the frequency $f (\mbox{Hz})$ with fix conformal time $|\eta| = 10^{-5}$ and small values of $\omega$ in both frames.}
\label{espec02}
\end{figure}

In this work we considered the quantum gravitational waves production during inflation in the BDT in the
         conformally related JBDF and EF. We saw that,
         in contrast with the GR, not only the scale factor but also the Brans-Dicke scalar field models the 
         time evolution of the cosmological gravitational waves in both frames. In the JBDF the scalar field appears in the gravitational
         waves equation and in the EF the scalar field changes the time evolution of the background.
    \par
         We computed the quantum particles number $N_k$ created during inflation and we saw that the number predicted by this model in the 
         JBDF approximates that predicted by GR when the Brans-Dicke parameter $\omega$ is big enough (depending on the 
         frequency, as we can see in Figures~\ref{num01}-\ref{num02}). The approximation between the EF and GR is achieved when the parameter 
         $\omega$ is negative and big.
    \par
         Afterwards we calculate three observable parameters: the quantum power spectrum $P_k$, the quantum spectral index $n_T$ and the quantum 
         energy density $\Omega(k,\eta)$ of the gravitational waves in both frames. The result obtained for the power spectrum $P_k$ shows that,
         different from the local observations and from the previous result acquired for the graviton number $N_k$, this model in the JBDF
         approximates the GR result one if $\omega$ assumes values  not so big, like $\omega=10$, at least for high frequency. For
         low frequency higher values of the Brans-Dicke are preffered, like $\omega=10^4$, as can be seen in Figure~\ref{espec01}. For $\omega>10^4$ the
         curve moves away from the GR one for all frequencies. In the EF we can see (Figure~\ref{espec02}) that this model approximates
         GR for small values of $\omega$, specially for high frequency. 
    \par
         For the spectral index we obtain a different result. We can compare the outcome of the GR for the spectral index $n_T$ with
         that of the BDT in the JBDF when $\omega>>1$, result that agrees with the one obtained for the quantum particle number
         $N_k$, and when $\omega\rightarrow\infty$ it yields a scale invariant perturbation. In the EF we can get scale invariant perturbation
         too, but for both limits $\omega\rightarrow\pm\infty$.
    \par
         In the case of the quantum energy density big values of $\omega$ make the energy density to explode and the GR prediction is
         approximated when $\omega$ is of the order of unity in the JBDF and in the EF $\omega<0$ but not $\omega<-20$ is preffered, 
         depending on the values of the frequency and the conformal time. 
    \par     
         The results obtained here suggest that the coupling constant $\omega$ of the BDT can vary with the cosmic scale \cite{omega01,omega02} and it
         depends substantially on the chosen frame. When gravitational waves are detected we can put some constraints on $\omega$ for large scales.
    \par
        Until there we can compute the scalar perturbations during inflation using the BDT, compare the results with these from the GR and, with the
         available observational data, calculate de ratio $r_{ST}$ between the tensor and scalar power spectra. The ratio $r_{ST}$ is related with the 
         power spectrum of the polarization modes of the CMB and, combined with the spectral index of the scalar perturbation, it allows us to rule in
         or out our inflationary model \cite{wmap5,wmap7}. We can also perform these same calculations but using from the beginning a background metric that is not flat but with a curvature, i.e., $k\neq 0$. Other possibility is to do the same for both scalar and tensor perturbation but in a model where the Brans-Dicke parameter $\omega$ varies with the scalar field $\phi$, \textit{i. e.}, $\omega=\omega(\phi)$. These calculations are in current investigations and will be presented elsewhere.

{\bf Acknowledgements:} We thank CNPq (Brazil) and CAPES (Brazil) for partial financial support. The authors would like to thank Clisthenis Ponce Constantinidis for helpful discussions. The suggestions made by the anonymous referee has permitted us to improve the work.



\begin{thebibliography}{99}


\bibitem{paul1} R. Crittenden, R.L. Davis and P.J. Steinhardt, Astrophys. J. {\bf 417}, L13 (1993), [arXiv:astro-ph/9306027];
\bibitem{paul2} R. Crittenden, J.R. Bond, R.L. Davis, G. Efstathiou and P.J. Steinhardt,Phys. Rev. Lett. {\bf 71}, 324 (1993), [arXiv:astro-ph/9303014];
\bibitem{mass} M. Giovannini, PMC Phys. {\bf A} 4, 1 (2010), [arXiv:0901.3026v1]; 
\bibitem{gwobs} LIGO: www.ligo.caltech.edu/\quad, \\ 
                VIRGO: www.virgo.infn.it/\quad, \\	
               	MiniGrail: www.minigrail.nl/\quad, \\
	              CLIO: www.icrr.u-tokyo.ac.jp\quad, \\
                GEO 600: www.geo600.org/\quad, \\
	              TAMA 300: tamago.mtk.nao.ac.jp/\quad, \\
	              Gr\'aviton Project: www.das.inpe.br/graviton/index.html\quad, \\	
	              AIGO: www.gravity.uwa.edu.au/\quad, \\
	              LCGT: gw.icrr.u-tokyo.ac.jp/lcgt/\quad, \\	
	              LISA: lisa.nasa.gov/\quad;
\bibitem{uzan} A. Riazuelo and J.P. Uzan, Phys. Rev. {\bf D62}, 083506 (2000), [arXiv:astr-ph/0004156];
\bibitem{ungarelli} A. Buonanno, M. Maggiore and C. Ungarelli, Phys. Rev. {\bf D55}, 3330 (1997), [arXiv:gr-qc/9605072];
\bibitem{sanchez} M.P. Infante and N. S\'anchez, Phys. Rev. {\bf D61}, 083515 (2000), [arXiv:hep-th/9907185];
\bibitem{gasperini} M. Gasperini, Phys. Rev. {\bf D56}, 4815 (1997), [arXiv:gr-qc/9704045];
\bibitem{riess} A.G. Riess et al, Astron. J. {\bf 116}, 1009 (1998), [arXiv:astr-ph/9805201];
\bibitem{quinte} R.R. Caldwell, R. Dave and P. J. Steinhardt, Phys. Rev. Lett. {\bf 80}, 1582-1585 (1998), [arXiv:astr-ph/9708069];
\bibitem{k} C. Armendariz-Picon, V. Mukhanov and P. J. Steinhardt, Phys.Rev.Lett. {\bf 85}, 4438-4441 (2000), [arXiv:astr-ph/0004134];
\bibitem{fantasma} V. Faraoni, Phys.Rev. {\bf D69}, 123520 (2004), [arXiv:gr-qc/0404078];
\bibitem{chap} N. Ogawa, Phys.Rev. {\bf D62}, 085023 (2000), [arXiv:hep-th/0003288]; J.C. Fabris, S.V.B. Goncalves and P.E. de Souza, Gen.Rel.Grav. {\bf 34}, 53-63 (2002), [arXiv:astr-ph/0203441]; M. C. Bento, O. Bertolami and A. A. Sen, Phys.Rev. {\bf D66}, 043507 (2002), [arXiv:gr-qc/0202064];
\bibitem{rose} J.C. Fabris, S.V.B. Gon\c calves and R. de Sa Ribeiro, Grav. Cosmol. {\bf 12}, 49-54 (2006), [arXiv:astro-ph/0510779];
\bibitem {ext} S. Capozziello and M. De Laurenti, Physics Reports, \textbf{509}, 167 (2011) [arXiv:1108.6266v2];
\bibitem{1} C. Brans and R. H. Dicke, Phys. Rev. {\bf 124}, 925 (1961);
\bibitem{2} P.A.M. Dirac, Proc. Roy. Soc. (London) {\bf A165}, 199 (1938);
\bibitem{3} P. Jordan, Z. Physik {\bf 157}, 112 (1959);
\bibitem{4} G.C. McVittie, M.N.R.A. Soc. {\bf 183}, 749 (1978);
\bibitem{barrow} J. D. Barrow, J. P. Mimoso and M. R. G. Maia, Phys. Rev. {\bf D48}, 3630 (1993);
\bibitem{dicke}  R. H. Dicke, Phys. Rev. {\bf 6}, 2163 (1962);
\bibitem{faraoni} V. Faraoni, E. Gunzig and P. Nardone, Fund. Cosmic Phys. {\bf 20}, 12 (1999), [arXiv:gr-qc/9811047v1];
\bibitem{santiago} D.I. Santiago and A.S. Silbergleit, Gen. Rel. Grav. {\bf 32}, 565 (2000), [arXiv:gr-qc/9904003v1];
\bibitem{scharre} P.D. Scharre, C.M. Will, Phys.Rev. {\bf D65}, 042002 (2002), [ arXiv:gr-qc/0109044];
\bibitem{plinio} J.P. Baptista,   J.C. Fabris and S.V.B. Gon\c calves, Astrophysics and Space Science {\bf 246}, 315-331 (1996), [arXiv:gr-qc/9603015];
\bibitem{Omega} D. Larson {\it et al}, Astrophys. J. Suppl. {\bf 192}, 16 (2011), [ arXiv:1001.4635];
\bibitem{dirac} P. A. M. Dirac, Phys. Rev. {\bf 114}, 924 (1959);
\bibitem{birrel} N. D. Birrel and P. C. W. Davies, {\it Quantum fields in curved space}, Cambridge University Press, Cambridge (1982);
\bibitem{inflacao} D. Baumann, \textit{TASI Lectures on Inflation}, Lectures from the 2009 Theoretical Advanced Study Institute at Univ. of Colorado, Boulder, [arXiv:0907.5424v1];
\bibitem{abramowitz} M. Abramowitz e I. A. Stegun, \textit{Handbook of Mathematical Functions}, National Bureau of Standards, (1964);
\bibitem{dodelson} S. Dodelson, \textit{Modern Cosmology}, Academic Press, (2003);
\bibitem{dicke01} R.H. Dicke, Phys.Rev. {\bf 125}, 2163 (1962);
\bibitem{omega01} M. Reuter and H. Weyer, Phys. Rev. {\bf D70}, 124028 (2004), [arXiv:hep-th/0410117];
\bibitem{omega02} I.L. Shapiro, J. Sola and H. Stefancic, JCAP {\bf 0501}, 012 (2005), [arXiv:hep-ph/0410095];
\bibitem{wmap5} E. Komatsu \textit{et al.}, Astrophys. J. Suppl. {\bf 180}, 330-376 (2009), [arXiv:0803.0547v2];
\bibitem{wmap7} E. Komatsu \textit{et al.}, Astrophys. J. Suppl. {\bf 192}, 18 (2011), [arXiv:1001.4538v3];


\end{thebibliography}
\end{document}